\documentclass{aa} 
\usepackage{graphicx}
\usepackage{txfonts}
\usepackage{ulem}
\usepackage{color}

\bibpunct{(}{)}{;}{a}{}{,} % to follow the A&A style
\begin{document}

   \title{Asymmetric shocks in $\chi$ Cyg observed with linear spectropolarimetry\thanks{Based on observations obtained at the T\'elescope Bernard Lyot (TBL) at Observatoire du Pic du Midi, CNRS/INSU and Universit\'e de Toulouse, France.}}

   \author{{\sc A.~L{\'o}pez Ariste}\inst{1},  {\sc B. Tessore}\inst{2}, {\sc E.S.  Carl\'{\i}n}, \inst{3,5,6} {\sc Ph.Mathias}\inst{1}, {\sc A. L\`ebre}\inst{4},  {\sc J. Morin}\inst{4},  {\sc P. Petit}\inst{1}, {\sc M. Auri\`ere}\inst{1}, {\sc D. Gillet}\inst{7}, {\sc F. Herpin}\inst{8}. 
   }
      \institute{IRAP, Universit\'e de Toulouse, CNRS, CNES, UPS.  14, Av. E. Belin. 31400 Toulouse, France \and
%\institute{IRAP - CNRS UMR 5277. 14, Av. E. Belin. 31400 Toulouse. France 
  %  \and Universit\'e de Toulouse, UPS-OMP, Institut de Recherche en Astrophysique et Plan\'etologie, Toulouse, France}
  Universit\'e Grenoble Alpes, CNRS, IPAG, 38000 Grenoble, France \and
   Instituto de Astrofisica de Canarias, E-38205, La Laguna, Tenerife, Spain  \and
LUPM, Universit\'e de Montpellier, CNRS, Place Eug\`ene Bataillon, 34095 Montpellier, France \and
Universidad de La Laguna, Dpto. Astrofisica, E-38206, La Laguna, Tenerife, Spain \and
Istituto Ricerche Solari Locarno, 6600, Locarno, Switzerland \and
Observatoire de Haute-Provence -- CNRS/PYTHEAS/Universit\'e
d'Aix-Marseille, 04870 Saint-Michel l'Observatoire, France  \and
Laboratoire d'astrophysique de Bordeaux, Univ. Bordeaux, CNRS, B18N, all\'ee Geoffroy Saint-Hilaire, 33615 Pessac, France. }

   \date{Received ...; accepted ...}

% \abstract{}{}{}{}{} 
% 5 {} token are mandatory
 
  \abstract
  {}
   {From a coherent interpretation of the  linear polarisation detected in the spectral lines of the Mira star $\chi$ Cyg, we derive information about the dynamics of the stellar 
   photosphere, including pulsation.}
    {From spectropolarimetric observations of $\chi$ Cyg, we perform careful analysis of the polarisation signals observed in atomic and molecular lines, both in absorption and emission, using the radiative transfer in the polarisation 
   context, through two mechanisms: intrinsic polarisation and continuum depolarisation.
   We also explain the observed line doubling   phenomenon in terms of an expanding shell in spherical geometry, which allows us to pinpoint the coordinates over the stellar disk 
   with enhanced polarisation.}
   {We find that the polarised spectrum of $\chi$ Cyg is dominated by intrinsic polarisation, with a negligible continuum depolarisation. 
   The observed polarised signals can only be explained by assuming that this polarisation is locally enhanced by velocity fields. 
   During the pulsation, radial velocities are not homogeneous over the disk. 
   We map these regions of enhanced velocities. }
   {We have set an algorithm to distinguish in any stellar spectra of linear polarisation the origin of this polarisation and the way to increase signal by coherently 
   adding many lines with an appropriated weight. 
   Applied to the Mira star   $\chi$ Cyg, we reached the unexpected result that during the pulsation, velocities are radial but not homogeneous over the disk. 
   The reason for these local velocity enhancements are probably related to the interplay between the atmospheric pulsation dynamics and the underlying stellar convection.}

%\cite{Wagenblast:1983aa}

   \titlerunning{Asymmetric shocks in $\chi$ Cyg from linear spectropolarimetry}
	\authorrunning{A. L{\'o}pez Ariste, B. Tessore,  et al.}
   \maketitle
%
%\cite{Bertout:1987aa}
\section{Introduction}

The recent discovery of strong linear polarisation signals in atomic lines of cool  and evolved stars has spurred an interest in interpreting these signatures and exploit them to 
learn about  photospheric conditions of these stars using new diagnostics.  
Betelgeuse provided the first case of study of this kind: \cite{auriere_discovery_2016} described those signals, and interpreted them as due to the depolarisation of the continuum.   The fact that some signal remains after integration over the stellar disk,  implies that photospheric (brightness) inhomogeneities  are present.
From this interpretation, \cite{lopez_ariste_convective_2018} were able to map this brightness distribution over the stellar disk, and to relate this distribution to the
 supergranulation of Betelgeuse.
The reconstructed images obtained with this new technique applied to spectropolarimetric observations  were corroborated by comparison with interferometric  observations. 
Over 4 years, the granulation of Betelgeuse has  been imaged, its spatial and temporal scales measured, and the velocities associated to this plasma estimated, 
altogether confirming that it is convection, granulation, that is at the origin of these signals \citep{Mathias:2018aa}. 
The technique has been extended by now to 2   more red supergiants: CE Tau and $\mu$ Cep ( Tessore et. al., 2019, in preparation).

In an attempt to generalise this sucessful result, other  types of cool and evolved stars have been considered  for similar linear polarisation signals   associated to spectral lines. One of them is the Mira star $\chi$ Cyg. %\textbf{(spectral type S6-9/1-2e). }
Clear linear polarisation signals had  indeed already  been detected in the Balmer lines of  Mira stars \citep[e.g. in the prototypical Mira star  $o$ Ceti; see ][]{fabas_shock-induced_2011}  , but, more important for  the present work, polarisation signals  were also detected in many other atomic and molecular lines over the spectrum of $\chi$ Cyg \citep{Lebre:2015aa}{

The interpretation of these linear polarimetric signals  associated to spectral lines is the purpose of this work.  In Section 2, we present the Mira star $\chi$ Cyg and all the observational material involved in this work. In Section 3, we investigate the origin of the linear polarisation associated to spectral lines. 
Starting from the signatures within the line profiles that suggest intrinsic polarisation, we have developed several tests to confirm this hypothesis.
This  kind of approach, with tests confirming initial hypothesis, is also used to interpret the nature of the surface inhomogeneities, necessary for
the polarisation signal to survive its integration over the stellar disk,  pointing in particular to local velocity increases. Our  main result is that the pulsation  of the Mira star $\chi$ Cyg, while being radial in direction, presents different velocities at different places over the disk.   In Section 5, we exploit the data from the spectropolarimetric survey we have performed on $\chi$ Cyg along 2015 and 2016, to map, with great caution, the places where these velocity enhancements appear to be. We speculate,  in Section 6, that these velocity enhancements may be the result of the interaction of the pulsation mechanism with either the general convective patterns or the returning material from previous cycles.

\section{Spectropolarimetry of the Mira star $\chi$ Cyg}

\subsection{Mira star $\chi$ Cyg}

Mira variables are low- to intermediate-mass asymptotic giant branch (AGB) stars that pulsate with a period of about one year. 
$\chi$\,Cyg is an S-type Mira star of spectral type S6-9/1-2e and with a pulsation period of about 408\,d.
High-resolution spectroscopic studies of Mira stars \citep{gillet_shock-induced_1983,alvarez_envelope_2000,alvarez_envelope_2001}  have reported strong emissions of the hydrogen lines lasting up to 80\,\% 
of the luminosity period. 
Those works have established that radiative and hypersonic shock waves, which are triggered by the pulsation mechanism, were periodically propagating throughout 
the stellar atmosphere, generating emission lines formation process and favoring the doubling of metallic lines. 
Well beyond the region the atmospheric regions probed with our data, the stellar neighbourhood of Mira stars has also been thoroughly studied through obser- vations of SiO masers or CN emissions  \citep[e.g.][]{Herpin:2006aa,Duthu:2017aa} which allowed to estimate the magnetic field strength in the circumstellar envelope, i.e. up to 8.8 Gauss in $\chi$ Cyg at a few stellar radii.  \cite{Vlemmings:2017aa} observed magnetically aligned dust and SiO maser polarisation in the envelope of the red supergiant VY Canis Majoris. All these observations are consistent with a toroidal field configuration in these objects.

From full Stokes spectropolarimetric observations, 
\cite {fabas_shock-induced_2011}  characterized the shock wave propagation throughout the stellar atmosphere of the prototypical oxygen-rich Mira star: $o$\,Ceti. 
They reported signatures in Stokes $Q$ \& $U$ but also in Stokes $V$ parameters (tracing linear and circular polarisation, respectively), 
associated to the strong Balmer hydrogen emissions known to be formed in the radiative wake of the shock wave \citep{fadeyev_structure_2004}. 
The origin of these spectropolarimetric signatures reveals a global asymmetry (at least partly photospheric) 
perhaps due to the passage of the shock's front throughout 
photospheric giant convective cells.   
%\sout{\color{red}  However, this explanation could not exclude the presence of a weak magnetic field  (that was still undetected at the surface of Mira stars at this epoch)}. 
 Few years later, \cite{lebre_search_2014} reported the first detection of a faint magnetic field at the surface of the S-type Mira $\chi$ Cyg, that is still to date the unique detection of a surface magnetic field for Mira stars.
\cite{Lebre:2015aa}  also reported for $\chi$\,Cyg} strong signatures in  Stokes $Q$ and $U$ profiles, associated to metallic lines. 
These features are strong, since they are detected from single observing sequences, and they are also variable along the pulsating phase. The  positions of these striking Stokes $Q$ and $U$ profiles,  was found well connected to the shock front position. 
%These linear polarization signatures reveal that during its propagation in the lower part of the atmosphere, the shock induces the radial direction on particles  as a peculiar geometry. 
 Moreover, \cite{Lebre:2015aa}   have also reported, from the Stokes U and Stokes Q spectra of $\chi$ Cyg, clear signatures associated to individual lines (e.g. SrI@460.7 nm, Na D2@588.9 nm). In the solar case, these peculiar lines are known to be easily polarizable in the presence of asymmetries at the photospheric level \citep[for a theoretical introduction and extensive bibliography of the solar case, the reader may refer for exemple to ][]{landi_deglinnocenti_polarization_2004}.

\subsection{Spectropolarimetric observations of $\chi$ Cyg}

\begin{table*}
\centering
\caption{Linear polarisation observations of $\chi$ Cyg (Stokes U and/or Stokes Q), from September 2007 to December 2016. For each observation the first column gives the observed Stokes parameter(s). The second and third columns give the date of observation in the Gregorian and Julian calendars, respectively. The fourth column gives the phase of the star ($\phi$), considering an ephemeris giving $\phi$ = 0 at maximum light  for JD=2457234.4 (30 July 2015) and a period of 408.7 days. All the observations  have been replaced on a single pseudo cycle (from $\phi$ = 0.00 to $\phi$ =1.00). The last columns give respectively, the exposure time (in seconds), the number of spectra for each parameter and the maximum signal-to-noise  ratio (S/N) in each spectrum. 
$^{a}$ observations that have been pushed together.
$^{b}$ One Stokes  parameter is missing for this date. Therefore it has not been considered in our study. }
\begin{tabular}{clccccc}
\hline
\hline
Stokes & Obs. date &  Julian date (JD) & phase  & exposure  & N  & S/N \\
& (dd/mm/yyyy)& (+2 450 000)& ($\phi$) & (in s) &(Q \& U) & (/2.6 ${\rm km.s^{-1}})$\\ 
\hline
Q \& U   &   04/09/2007   &   4347.86 & 0.94  &   400   &   1 \& 1   &   1527 \\ %0.95, 0.94
Q \& U   &   28/05/2015   &   7171.11 & 0.84 &   400   &   2 \& 2   &   1464 \\ %0.84
Q \& U   &   01/06/2015   &   7175.12 &0.85  &   400   &   2 \& 2   &   1488 \\ 
Q \& U   &   26/06/2015$^{a}$   &   7200.05 &0.91  &   400   &   1 \& 1  &   1369 \\ 
Q \& U   &   28/06/2015$^{a}$   &   7202.02 & 0.92  &   400   &   1 \& 1  &   1479 \\ 
Q \& U   &   19/07/2015   &   7222.98  & 0.97  &   400   &   3 \& 4   &   1391 \\ 
%Q \& U   &   05/08/2015$^{a}$   &   7239 & 0.01 &   400   &   4 \& 4   &   30 {\color{red} OUT ?}\\ 
%Q \& U   &   06/08/2015$^{a}$   &   7240& 0.01  &   400   &   4  \& 4   &   30 {\color{red} OUT ?}\\ 
Q \& U   &   10/08/2015   &   7245.01 &  0.02 &   400   &   4 \& 4   &   990 \\ 
Q \& U   &   20/08/2015   &   7254.91 & 0.05  &   400   &   4 \& 4   &   1270 \\ 
Q \& U   &   10/09/2015$^{a}$   &   7275.96  &0.10   &   400   &   2 \& 2   &   1470 \\ 
Q \& U   &   11/09/2015$^{a}$   &   7276.82 & 0.10  &   400   &   2 \& 2  &   1260 \\ 
Q \& U   &   09/10/2015   &   7304.93 &   0.17&   400   &   2 \& 2   &   1568 \\
Q \& U   &   16/10/2015   &   7311.80 &  0.19 &   400   &   2 \& 2   &   904 \\ 
Q \& U   &   15/11/2015$^{a}$   &   7341.86 &   0.26&   400   &   2 \& 2   &   882 \\ 
Q \& U   &   16/11/2015$^{a}$   &   7342.74 &  0.26 &   400   &   2 \& 2   &   1048 \\ 
%Q \& U   &   01/12/2015   &   7357.74 &  0.30 &   400   &   2 \& 2   &   1018 \\ 
%U   &   06/04/2016$^{b}$   &   7484 & 0.61 &   400   &   4   &   533 {\color{red} OUT ?}\\ 
Q \& U   &   07/08/2016$^{a}$   &   7607.88  &  0.91 &   400   &   4 \& 4   &   1349 \\ 
Q \& U   &   08/08/2016$^{a}$   &   7608.89 &0.92  &   400   &   4 \& 4   &   1347 \\ 
Q \& U   &   24/08/2016   &   7625.00  &   0.95&   400   &   4 \& 4   &   1556 \\ 
Q \& U   &   24/09/2016   &   7655.90  & 0.03  &   400   &   3 \& 3   &   1096 \\ 
Q \& U   &   21/10/2016   &   7682.83  &  0.09 &   400   &   3 \& 3   &   1183 \\ 
Q   &   26/10/2016$^{b}$   &   7687.86 & 0.11 &   400   &   3  &   909 \\ 
U   &   27/10/2016$^{b}$   &   7687.86 & 0.11  &   400   &   3  &   682 \\ 
Q \& U   &   07/12/2016   &   7729.73 &0.21  &   400   &   4 \& 4   &   1548 \\ 
Q \& U   &   20/12/2016   &   7742.75 & 0.24  &   400   &   4 \& 4   &   1224 \\ 
\hline
\end{tabular}
\label{annexeLP:chicyg}
\end{table*}

  Table \ref{annexeLP:chicyg} shows the series of observations of $\chi$ Cyg,  at our disposal for the present work. Except for the first observation (date of 4 September 2007)  performed during a previous observing campaign, all 2015-2016 observations have been obtained in the same Large Program conducted with the Narval spectropolarimetric instrument mounted at the Telescope Bernard Lyot (TBL, Pic du Midi, France). 

\begin{figure}
   \centering
   \includegraphics[width=0.5\textwidth]{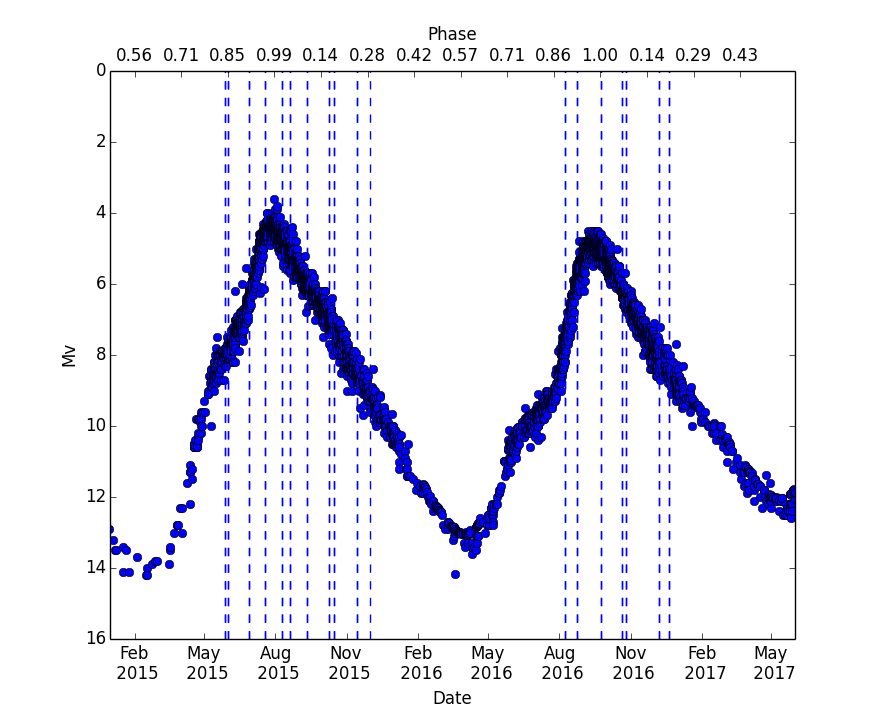}
      \caption{Visual magnitude of $\chi$\,Cyg as measured by AAVSO. The vertical dashed lines mark the dates for which spectropolarimetric data are available.}
         \label{AAVSO}
   \end{figure}

Figure\,\ref{AAVSO} shows the time series of the visual magnitude of $\chi$\,Cyg as measured by the AAVSO. 
The vertical dashed lines   indicate our available 2015-2016 observations, considering that some observations collected at close dates (or consecutive nights) have been combined into one observational phase (cf Table \ref{annexeLP:chicyg}).  All these observations appear to have been  collected around a maximum light ($\phi$ from  0.8 to 1.3), when  the atmospheric shock wave  is known to  have always an upward motion while propagating throughout the stellar atmosphere \citep{Gillet:1985aa}.
%%%%%%% agnes  : j'ai commente les  figures car je n'ai pas les fichiers, mais les captions sont OK)
The Stokes U and Stokes Q observations of September 4th, 2007 (also collected with Narval at TBL) show clear signatures in individual lines.  Since they were also the first data to be scrutinised, these 2007 observations will be further used to illustrate the origin of the linear polarisation associated to atomic and molecular lines (see Section 3). The 2015-2016 observations, resulting from a regular spectropolarimetric monitoring of  $\chi$ Cyg (on a monthly basis when the star was observable from Pic du Midi), will be used in Section 4 to map photospheric conditions.

%\section{Derivation of the model}
\section{On the origin of the linear polarisation associated to the spectral lines of $\chi$ Cyg}

polarisation signals in stellar atomic lines are weak, often below 0.1\% of the continuum intensity. 
It is customary, in stellar spectropolarimetry, to add the signals over many lines throughout the spectrum with the goal of increasing the S/N ratio. 
Such algorithms (e.g. Least Squares Deconvolution -- LSD --,\cite{donati_spectropolarimetric_1997,kochukhov_least-squares_2010}, SLA, \cite{paletou_critical_2012} and others) depend on the self-similarity assumption, according to which all lines to be co-added are formed at similar heights and carry a signal with a similar spectral shape up to a scale factor. This has been traditionally the case 
 of circular polarisation, related to the Zeeman effect    and used to detect surface magnetic fields.
 %, \sout{spectra are resampled in terms of velocity (to cancel the Zeeman effect over the whole line profile and 
%the spectrograph resolution),} one expects the typical S-profile to appear in all Zeeman-sensitive lines,  with an amplitude proportional to the Land\'e factor of each line.

The situation faced here with $\chi$ Cyg is that  the S/N ratios are barely enough to be reliable only through LSD technique. But trusting LSD implies that the signatures in individual lines  are similar, though the S/N in those lines is in general too low to check this assumption.
In order to justify the use of LSD to study linear polarisation of $\chi$\,Cyg, we started by an in-depth analysis of the signals  detected in key individual lines.

Among the observations of $\chi$\,Cyg   (presented in Table \ref{annexeLP:chicyg}), several observations stand out because of the large amplitude of the linear polarisation  signals over most of the visible spectrum. The largest amplitudes seen ( September 4th, 2007 or August 8th and 16th, 2007) coincide with the  dates around the maximum light (phases 0.91, 0.94 and 0.95).
We pick one of  such singular observations, the one on September 4th, 2007, close to the occurence of a maximum light ($\phi$ = 0.94). According to the shock propagation scenario from \cite{Gillet:1985aa}   this observation corresponds to the following pulsation state for a Mira star : a strong shock wave has already emerged from the photosphere (just before the maximum light, at around $\phi$ = 0.8) and it is propagating throughout the stellar atmosphere, until its final  fading around $\phi$ = 1.4. It is at this stage that the observation of September 4th, 2007 produced strong signals, sufficient to examine individual lines and reveal the physical mechanisms at work in the production of linearly polarised light.

\subsection{Radiative transfer equation}
Roughly speaking, linear polarisation in atomic lines of a stellar spectrum can have two origins: Zeeman effect or scattering. 
For  $\chi$ Cyg, \cite{lebre_search_2014} detected a very weak circular polarisation signature  around its maximum light of March 2012, revealing the presence of a  faint surface magnetic field ($2-3\,G$).
However, its effect on the linear polarisation is expected to be one order of magnitude smaller.
Therefore,  in the case of $\chi$ Cyg, a Zeeman contamination of the linear polarisation can be excluded. And thus we retain in the following only scattering processes.

In this context, within the stellar atmosphere, an atom has three different ways of creating a spectral polarisation signature: continuum
depolarisation through lines, intrinsic line polarisation and finally Rayleigh and/or Thomson scattering.
We follow \cite{landi_deglinnocenti_polarization_2004} to write, in the absence of magnetic fields, a radiative transfer equation for the Stokes parameter $Q$ defined as positive when 
the polarisation plane is perpendicular to the scattering plane\footnote{This plane is defined by the scattering point, the centre of the star and the observer.} 
of a particular point in the stellar atmosphere, taking advantage that this polarisation is always small:
\begin{eqnarray}
\frac{d}{ds}Q(\nu,\vec{\Omega})=-\left[k_{\nu}^c+k_L^A\phi(\nu_0-\nu)\right]Q(\nu,\vec{\Omega})+ \nonumber \\
\frac{3}{2\sqrt{2}}k_L^A\sin^2 \theta \left[w^{(2)}_{J_uJ_l}\sigma^2_0(J_u)S_L-w^{(2)}_{J_lJ_u}\sigma^2_0(J_l)I(\nu,\vec{\Omega})\right]\phi(\nu_0-\nu)+ \nonumber \\
\frac{3}{2\sqrt{2}}k_v^c\sin^2\theta\sum_{i=1}^3\beta_i\int d^3\vec{\rm v}_i f(\vec{\rm v}_i)\left[J^2_0\left(\nu-\nu\frac{\vec{\rm v}_i\cdot\vec{\Omega}}{c}\right)\right]_{\vec{v}_i}
\label{RTE}
\end{eqnarray}
where the 3 considered processes  quoted above  correspond to each line on the right-hand side, respectively.

The equation describes the evolution of the Stokes $Q$ parameter along a path $s$ as a function of frequency $\nu$ and in the direction $\vec{\Omega}$. 
An atomic line is assumed to be at $\nu_0$ in the proximity of the frequency $\nu$ with a characteristic line profile described by $\phi(\nu_0-\nu)$. 
In spite of being a first order approximation, this equation is quite complex at first sight. 
It unveils the richness of the scattering polarisation spectra of a star, even in the absence of magnetic fields. 
It is this richness that led \cite{stenflo_second_1997} to coin the name {\it second solar spectrum} to refer to this scattering polarisation spectrum in the case of the Sun. 
Any hope in using LSD techniques requires the identification of a large enough number of atomic lines that are sensitive to just one of those three processes explicited in the right-hand side of this equation. 

%{\color{red} NE FAUDRAIT il PAS JUSTIFIER QUE LE CHAMP MAG detecte dans chi Cyg n'empeche pas de considerer ce formalisme, prevu pour etre "en l'absence de champ mag"! ?}

\subsubsection{Rayleigh and/or Thompson scattering}
In the absence of lines  sensitive to the first 2 mechanisms in Eq.\ref{RTE}, the only non-zero term in the equation would be the last one, which describes Rayleigh and/or Thompson 
scattering of photons over electrons ($i=1$), H atoms ($i=2$) and He atoms ($i=3$). 
Scattering over heavier atoms can be safely neglected. 
Rayleigh scattering linearly polarizes light proportionally to $\sin^2 \theta$, the scattering angle. 
The amount of light is obviously proportional to the absorption coefficient of the continuum at that frequency $k_{\nu}^c$, but also to the fractional contribution 
$\beta_i$ of electrons, H and He atoms to that opacity. 
For a scatterer to emit net polarisation, it should be illuminated by an anisotropic radiation field described by the non-zero 2nd-rank spherical tensor of the radiation field 
$J^2_0$ (which in the first approximation can be written as $3K-J$ with $K $ and $J$, the 2nd and 0th order momenta of the angular distribution of the specific intensity) which depends on frequency in the rest-frame of the scatterer. 
An integral over the velocity distribution $f(\vec{\rm v}_i)$ is necessary to convert this dependency to the observer's frame.  
If the continuum forms over a large enough atmospheric layer for anisotropy to be non-negligible, one expects the continuum to be polarised. 
In the Sun this continuum polarisation amounts to 0.1\,\% \citep{leroy_new_1972,stenflo_polarization_2005}, In cool stars like Betelgeuse ($\alpha$\,Ori) it grows to 1\,\% of the continuum intensity \citep{clarke_polarization_1984,doherty_polarization_1986}. In the  Mira star $\chi$ Cyg, \cite{Boyle:1986aa}  reported (from polarimetric observations collected around a maximum light) that the polarisation of the continuum varies from 1\% in the blue part  down to 0.25\% in the red part of the spectrum. They also reported striking enhancements in the linear polarisation level associated to Balmer emission lines (+ $\sim$ 0.5\%), to  molecular bands (+$\sim$ 0.5\%), and to the \ion{Ca}{I} line at 422.6 nm (+$\sim$ 2\%) pointing to  a polarisation likely arising from the stellar atmosphere, in the  region of formation of  the spectral lines.

\subsubsection{Continuum depolarisation}
A spectral line that forms above a continuum polarised by Rayleigh scattering absorbs linearly polarised photons and re-emits {\it a priori} unpolarised photons. 
Spectral lines therefore tend to de-polarize the continuum. 
This process is described by the first term of Eq. \eqref{RTE} where this de-polarisation of the incoming polarisation $Q(\nu,\vec{\Omega})$ is described proportional 
to the combined absorption coefficients of the continuum and the line, $k_{\nu}^c+k_L^A$, times the spectral profile of the line. 
It depends on the details of the stellar atmosphere with respect to the formation regions of the continuum and every individual lines, to estimate the impact of this 
depolarisation mechanism on the second stellar spectrum. 
In the case of the Sun, well over 90\,\% of the spectral lines depolarize the continuum. 
In particular, with few exceptions, all the numerous \ion{Fe}{I}  lines depolarize the continuum. 
\cite{auriere_discovery_2016} found that, in the case of Betelgeuse, all atomic lines did depolarize the continuum and since the spectral shape of a line $\phi(\nu_0-\nu)$ 
can be considered similar enough to that of any other line, this allowed those authors to use LSD on the linearly polarised spectrum of Betelgeuse and to map the 
presence of bright spots on its photosphere.

\subsubsection{Intrinsic line polarisation }
%\footnote{Our use of the adjective {\it intrinsic} is at odds with its use by other cited authors and may lead to confusion. The presently accepted  explanation for the  intrinsic polarization of $D_1$, {\it intrinsic} in our present use, is that alignment in the lower  level is due  to a transfer  of  alignment from the upper level of $D_2$, and because of this those authors describe it  as {\it extrinsic} polarization of $D_1$ in contrast to the {\it intrinsic} polarization of $D_2$ in  which alignment  is induced exclusively by  anisotropic illumination and not by transfer from other aligned levels. In  our opinion, it is  the alignment in $D_1$ which  is {\it extrinsic}, while the polarization is still generated intrinsically by a transition between aligned levels. With this argument, we keep our use of the adjective {\it intrinsic} to refer to polarization arising from emission of aligned levels, whether this alignment is 
%{\it extrinsic} or {\it intrinsic} in origin.} }

Competing with depolarisation, one finds the second, and most complex, term of Eq.\eqref{RTE}. 
It describes what we refer to as the intrinsic polarisation of a line \footnote{In contrast to previous uses by other authors of the adjective {\it intrinsic} and at the risk of some confusion, we call {\it intrinsic}  any signal with origin in this term which does not explicitly depend on the incoming light or its polarisation as the two other terms. Evidently, the atomic polarisation appearing in this 2nd term may have its origin in anisotropic radiation, but as explained, there is a clear difference in the type of signals expected from one and the other terms in Eq. (1) hence justifying our use of {\it intrinsic}. }.
Anisotropic illumination can introduce both population imbalances or coherences among otherwise degenerated atomic sublevels. 
They are quantum in nature and in the present case reduce to  the $\sigma^2_0 (J)$ spherical tensor of the atomic density matrix corresponding to the level with total 
angular momentum $J$ , an imbalance of populations between sublevels that we call atomic alignment. 
Atomic alignment in the upper level results in polarisation in the emitted light, and this is why we multiply it by the source function $S_L$. 
Atomic alignment in the lower level can be seen as absorbed incoming photons, and this is why we multiply it by the incoming intensity $I(\nu,\vec{\Omega})$. 
Not all transitions between atomic levels are equally efficient in producing polarisation in the presence of atomic alignment:
the $w^{(2)}_{J_uJ_l}$ quantum  coefficient expresses this efficiency as a function of the upper and lower total angular momenta. 
Its explicit definition can be found for instance in  \cite{landi_deglinnocenti_polarization_2004}. Particular values of the upper and lower angular momenta cancel this coefficient. For example transitions from ($J_l=$) $\frac{1}{2}$  to ($J_u=$) $\frac{1}{2}$ , or from 1 to 0, 
or from $\frac{3}{2}$ to $\frac{1}{2}$, have all $w^{(2)}_{J_uJ_l}=0$. All those atomic lines in the stellar spectra with those values of the total angular momentum for their lower and upper levels will be unable to produce intrinsic polarisation in consequence, any polarisation signal in such selected lines may only be attributed to depolarisation. In the presence of noise, these conclusions can be extended to other case with small absolute values of $w^{(2)}_{J_uJ_l}$.

This  provides one possible way to distinguish the origin of observed linear polarisation in a stellar spectrum. 
Following \cite{auriere_discovery_2016} we consider the Na\,$D$ doublet. 
Neglecting hyperfine structure, the $D_1$ line arises from a $\frac{1}{2}-\frac{1}{2}$ transition with $w^{(2)}_{\frac{1}{2},\frac{1}{2}}=0$. 
This means that the $D_1$ line {\em cannot} produce intrinsic polarisation.  
Note that the hyperfine structure modifies the past assertion: among the several hyperfine transitions that form the $D_1$ line, some can indeed carry intrinsic polarisation.
This can be seen in the second solar spectrum, where the $D_1$ line indeed shows a signal \citep{casini_atomic_2002,stenflo_second_1997,trujillo_bueno_physical_2002} .
But this is a signal at least one order of magnitude smaller than that present in $D_2$. 
Indeed, $D_2$ has $w^{(2)}_{\frac{3}{2},\frac{1}{2}}=0.6$ and it is prone to show large intrinsic signals. 
Both $D_1$ and $D_2$ components are, on the other hand, equally efficient in depolarizing the continuum, with almost identical conditions of formation and frequency. 

By looking into the second spectrum of Na\,$D$ lines, one can therefore identify two extremes: if $D_1$ shows a polarisation signal with a similar amplitude than that of $D_2$,
then the lines are depolarizing the continuum. 
On the other hand, if the $D_1$ linear polarisation is much smaller than that of $D_2$, then the signals are dominated by intrinsic polarisation.  

The first scenario was found by  \cite{auriere_discovery_2016} on Betelgeuse: both $D_1$ and $D_2$ components did show similar amplitudes, and the interpretation was that the atmospheric 
structure in that star leads to a depolarisation of the continuum (first term of Eq.\eqref{RTE}). 
Indeed, if $D_2$, a line so prone to show strong intrinsic signals, is depolarizing continuum, one can safely assume that so do all the lines in the spectrum. 
The Sun itself shows a mixed scenario: 
while most of the lines just depolarize the continuum, a few lines form in conditions adequate to show intrinsic polarisation, $D_2$ prime among them.

\subsection{Dominating process for $\chi$\,Cyg}

\begin{figure*}
   \centering
   \includegraphics[width=0.45\textwidth]{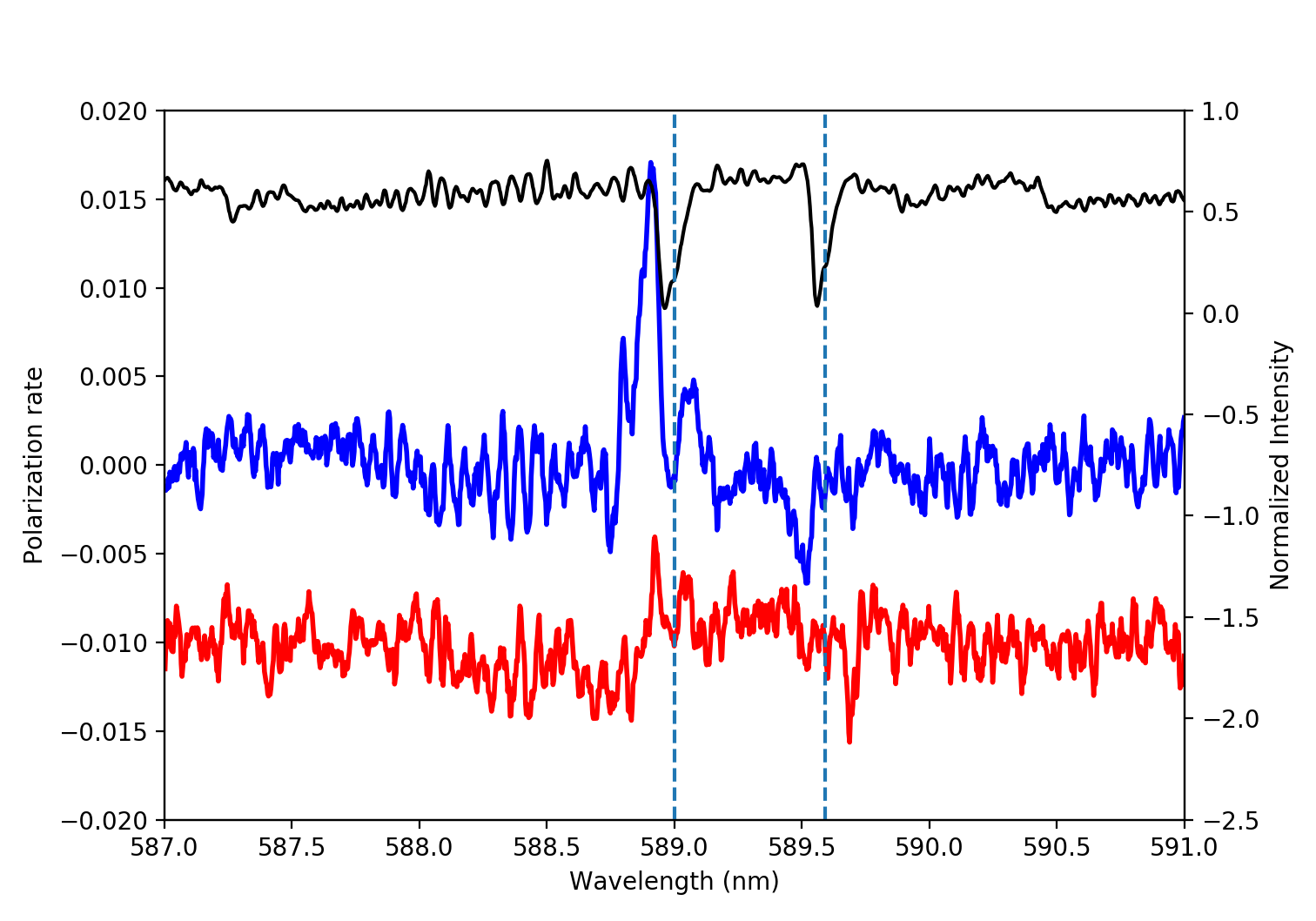}
\includegraphics[width=0.45\textwidth]{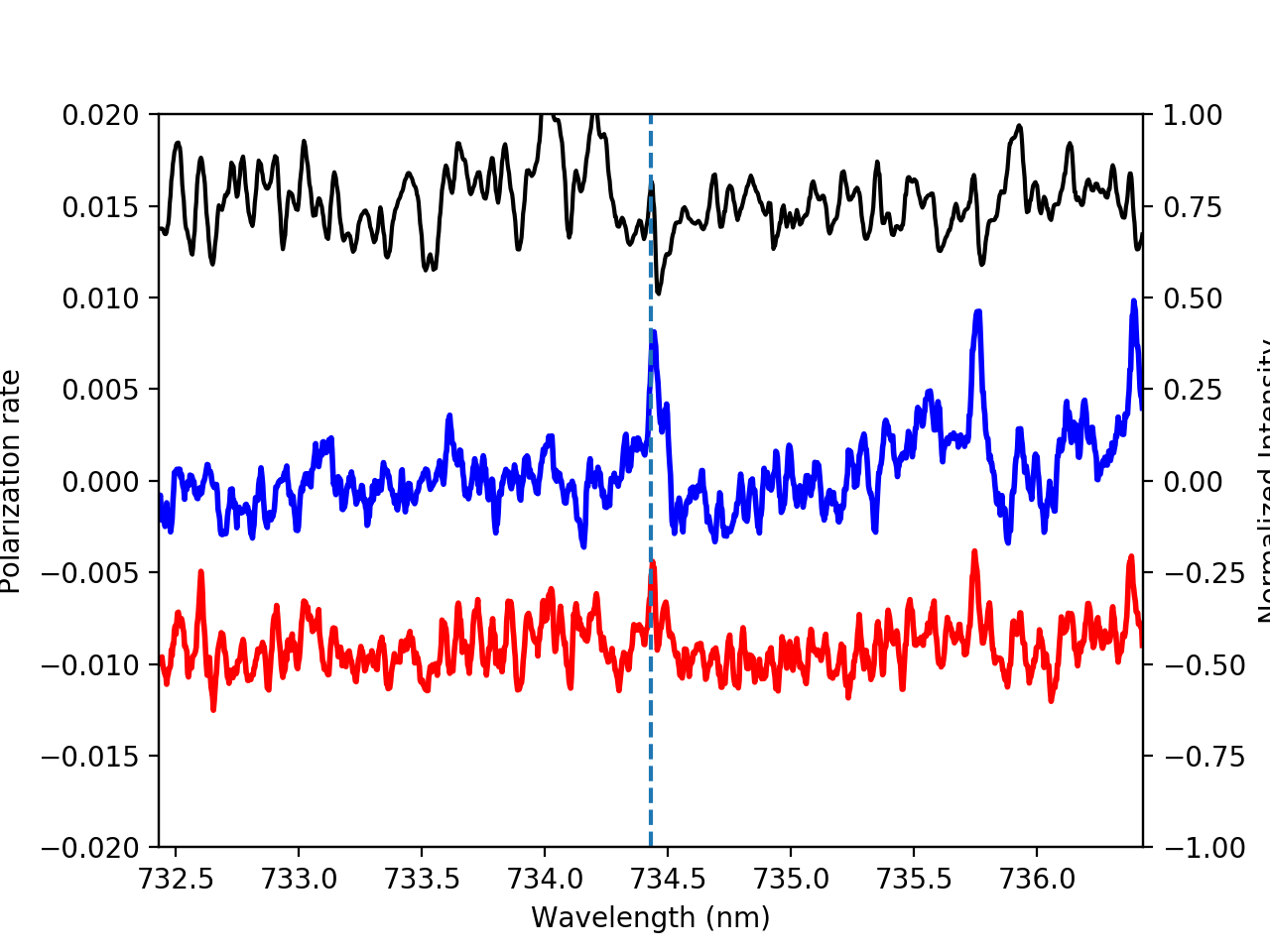}
   
	  \caption{Spectra of $\chi$\,Cyg on September 4th, 2007. Left: around the region of the Na\,$D$ doublet; Right: around a \ion{Ti}{i} line in the near infrared spectrum. 
	  The thin black line shows the intensity (ordinate axis on the right) in normalized units, whereas the thick coloured lines (blue for $U$, red for $Q$) show the 
	  linear polarisation spectra.
	  The rest wavelength of both Na\,$D_1$ and $D_2$ in the left plot, and of the \ion{Ti}{i} line in the right one, are represented with a vertical dashed line.}
	  
         \label{NaD}
   \end{figure*}

We now consider the Na\,$D$ doublet in the spectrum of $\chi$\,Cyg obtained on September 4th, 2007 (Fig.\,\ref{NaD}).  
Note that even considering a night with one of  the largest linear polarimetric signals, the amplitude of the polarisation signature is  weak.
This is partly due to a low S/N ratio in individual lines, but also for the ubiquitous presence of molecular lines (see below). 
In spite of those problems, one can identify a double peak around $D_2$, positive in both $Q$ and $U$,
whereas nothing is present in the same wavelength range around $D_1$. Small peaks with opposite amplitude are seen right beyond the red wing of $D_1$ in $Q$ and beyond the blue wing in $U$. The coherence of polarisation signs in all absorption lines that we are going to uncover in the text below, and the absence of any spectral feature at this wavelength, led us to ignore these signals (the 2nd solar spectrum shows a small signal in $D_1$, one order of magnitude smaller than the signal in $D_2$, but it remains difficult to relate it to these peaks seen in our data).
 With this caveat in mind, we find in this absence of signal in $D_1$ a first indication that intrinsic polarisation dominates the Na\,$D$ lines for $\chi$\,Cyg.

%%%%%%% agnes  : j'ai commente les  figures car je n'ai pas les fichiers, mais les captions sont OK)

Also, Fig.\,\ref{NaD} (right) shows, for the line \ion{Ti}{i} at 734.4\,nm, signals in both $Q$ and $U$ signals that are of the same positive sign as those present in the Na $D_2$ component. 
It is not the only signature with identical sign in that region, and around 736\,nm another line (probably \ion{Ti}{i} at 735.774nm) also shows this similar signature.  Let us recall that while most lines present a single gaussian-like polarisation profile, the  Na $D_2$ line presents a characteristic multi-peak profile due to its rich fine atomic structure. The comparison between this and  lines like the \ion{Ti}{i} in this figure must be limited to the signs of the $Q$ and $U$ profiles.

%{\bf on peut aussi remarquer que la structure en double pic est bien plus large pour NaD que pour TiI: effet de hauteur de formation? Ca peut aussi montrer que le profil moyen est... tres moyenne... faire des sous-masques pour avoir une figure montrant la structure en double-pic a partir de la hauteur de formation?.
%ALA:  Le double pic est une propriete assez unique de la raie D2. Il y a tres peu de raies avec cette double structure. Ce qui est normal c'est un signal similaire au profil d'intensite comme celui du Ti, ou de la raie moyenne} {\color{red} {\bf Oui ! moi je ne suis même pas convaincue qu'il y ait de la structure en double pic dans les signatures des deux raies de TiI. En tous cas rien de comparable à la structure double Pic de NaD2. Autant passer toute cela sous silence.}}\\
A trend appears here: selected lines show a polarisation signature which happens to be positive in both $Q$ and $U$ for the date of September 4th, 2007 and,  in the case of the Na\,$D$ lines, this has to be interpreted as intrinsic polarisation. 
The set of lines showing analogous signatures is much larger but it quickly gets messed with the many molecular lines and bands in the spectrum of this cool star. 
This is the reason for having selected, for the plot, this  otherwise unremarkable \ion{Ti}{i} line  around $736\,nm$: it is  located in a region where molecules appear to not overwhelm the spectrum. 

\begin{figure*}
   \centering
   \includegraphics[width=\textwidth]{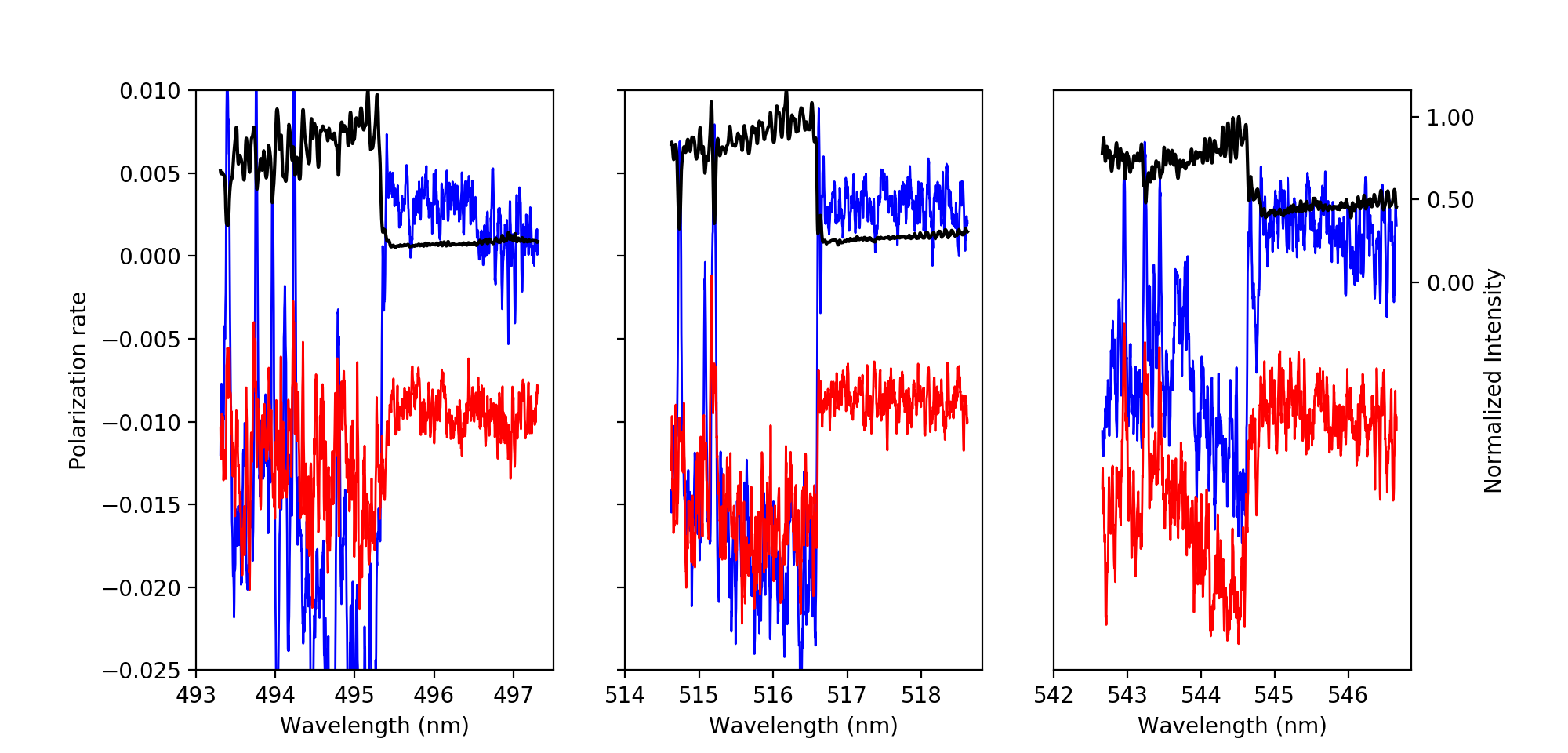}
      \caption{Spectra of $\chi$\,Cyg on September 4th, 2007 in three spectral regions where molecular bandheads of TiO  and VO are found. As in Fig.\,\ref{NaD}, the black thin line represents intensity 
	  (ordinates at right) and Stokes $Q$ and $U$ are shown in red and blue respectively. $Q$ has been shifted 0.01 units for clarity.}
         \label{mol}
   \end{figure*}

Molecular scattering polarisation is a much richer world than what we have explored until now \citep{asensio_ramos_evidence_2005,asensio_ramos_theory_2006,landi_deglinnocenti_polarization_2004}. 
But in general, molecules do present alignment and intrinsic line polarisation in an analogous manner to atoms \citep{landi_deglinnocenti_polarization_2004}. 
Are the $Q$ and $U$ signals common to both Na\,$D_2$ and \ion{Ti}{i} lines presented in Fig.\,\ref{NaD} also present in molecular lines?
Fig.\,\ref{mol} shows three different molecular bandheads along the visible spectrum of $\chi$\,Cyg. 
Molecular transitions cluster around the bandhead and concentrate the signal: the intensity goes down and linear polarisation shows also a drastic positive jump both in $Q$ and $U$.

%%%%%%% agnes  : j'ai commente les  figures car je n'ai pas les fichiers, mais les captions sont OK)

Further examination of other molecular band heads and individual molecular lines confirm that a net linear polarisation signal positive both in $Q$ and $U$ is ubiquitous 
in molecular as  well as  in atomic lines. 
 However, this signal is present in the Na\,$D_2$ line, but not in the Na\,$D_1$ line. Hence, we make the hypothesis that \textbf{all} these atomic and molecular lines are showing intrinsic line polarisation through scattering in an aligned atom or molecule. 
Taking this hypothesis to its extreme, we cancel in Eq.\eqref{RTE} the continuum polarisation (3rd) and the line depolarisation (1st) terms. 
Consequently, Eq. \eqref{RTE} becomes:
\begin{equation}
Q=\frac{3}{2\sqrt{2}}k_L^A\sin^2 \theta \left[w^{(2)}_{J_uJ_l}\sigma^2_0(J_u)S_L-w^{(2)}_{J_lJ_u}\sigma^2_0(J_l)I(\nu,\vec{\Omega})\right]\phi(\nu_0-\nu)
\label{TE_intrinsic}
\end{equation}
The first remark about this equation is that all the spectral shape dependence around a given spectral line is in the generalised profile $\phi(\nu_0-\nu)$. 
This points toward the possibility of adding up lines through LSD \citep{donati_spectropolarimetric_1997} after appropriate rescaling of the frequency. 
Of course,  in this case, the weighting of the individual lines cannot be the Land\'e factor   \citep[the usual weighting in Zeeman Doppler Imaging; see][]{donati_spectropolarimetric_1997}.
Inspection of Eq. \eqref{TE_intrinsic} allows further simplification by noticing that, more often than not, we can assume that the lower level of most transitions, 
because of its stability, will probably be depolarised by collisions and therefore $\sigma^2_0(J_l)=0$. 
We also notice that, in a two-level atom, the alignment of the upper level will be given by
\begin{equation}
\sigma^2_0(J_u)=w^{(2)}_{J_uJ_l}\frac{J^2_0}{J^0_0}
\label{sigma20}
\end{equation}
and therefore the emitted polarisation can be approximated by:
\begin{equation}
Q=\frac{3}{2\sqrt{2}}k_L^A\sin^2 \theta  (w^{(2)}_{J_uJ_l})^2 \frac{J^2_0}{J^0_0} \phi(\nu_0-\nu)
\label{TE_intsimp}
\end{equation}
From this latter expression, we realise that  the appropriate weighting factor for LSD is the coefficient $(w^{(2)}_{J_uJ_l})^2 $ which depends exclusively on the atomic numbers of the upper 
and lower levels. 

If the other 2 processes were also present, it can be noticed that, independently of the sign of $w^{(2)}_{J_uJ_l}$ (which can be either positive or negative depending on the 
quantum numbers of the involved levels), the emitted polarisation will always have the same  sign: perpendicular to the scattering plane. 
On the other hand, continuum polarisation is also perpendicular to this plane. Depolarisation will cancel this signal, and it will appear in our polarimeters as a signal of the opposite sign:  parallel to the scattering plane. 
Thus, intrinsic line polarisation and depolarisation processes will have a similar spectral shapes but opposite signs. 
If, contrary to our extreme assumption of all lines showing intrinsic line polarisation, a subset of lines shows a depolarisation signal (as  it is the case in the second solar spectrum) the polarisation signal would cancel out when summing up lines, and the LSD profile would end up with an amplitude much smaller than that seen in 
individual lines as the ones shown in Fig.\,\ref{NaD}. 
This provides us with an \textit{a posteriori} test upon our hypothesis (Section 4).

\subsection{polarisation signal and asymmetry}
\begin{figure*}
   \centering
   \includegraphics[width=\textwidth]{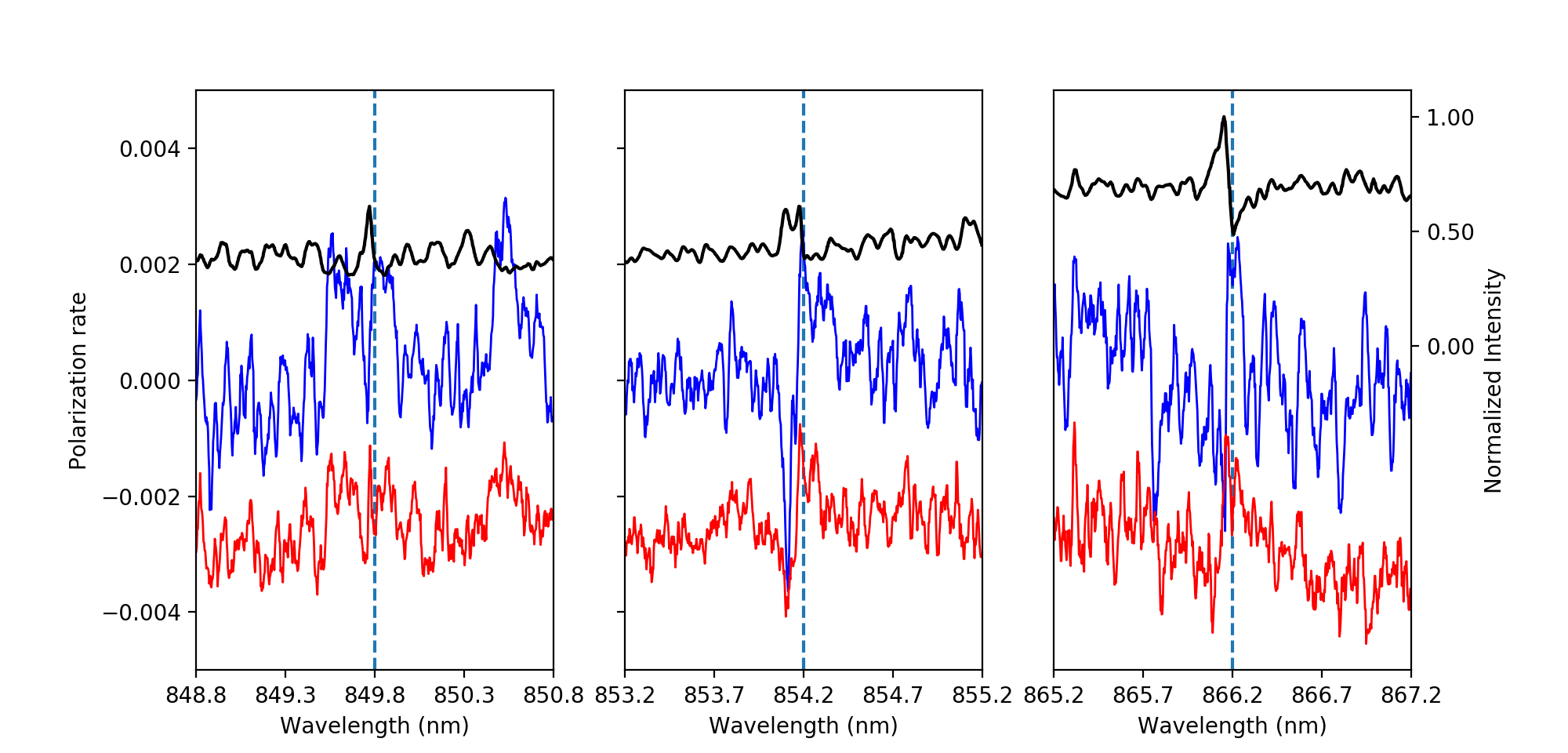}
      \caption{Spectra of $\chi$\,Cyg on September 4th, 2007 around the three lines of \ion{Ca}{II} triplet in the near IR. 
	  As in Fig.\,\ref{NaD}, the black thin line represents intensity (ordinates at right) and Stokes $Q$ and $U$ are shown in red and blue respectively.  $Q$ has been shifted 0.01 units for clarity.
	  The lines are split into a blue-shifted emission, and a red-shifted absorption. 
	  The polarisation changes sign accordingly as if dominated by the amplification of anisotropy in the presence of velocity gradients. }
         \label{CaIR}
   \end{figure*}

Polarised stellar spectra are the result of the integration of the signal over the stellar disk. 
We have found a source of local linear polarisation in the intrinsic polarisation of the lines formed by scattering. 
But since the scattering plane rotates with the position angle of the scattering point over the stellar disk, the signs of $Q$ and $U$ change around the stellar disk and, 
if the star is homogeneously polarizing its spectrum, the result is a zero net signal. 
Since {for $\chi$\,Cyg a net polarisation is observed, we must conclude that the stellar disk is not centrally symmetric with respect to polarisation. We must conclude that there is one region of the stellar disk with a larger emission of polarised light.
%Even if the observed polarization is the residual after most of the signal has cancelled out, it is interesting to determine the source(s) of inhomogeneities.

There are two manners in which one particular region over the stellar disk can dominate the integral which computes the net polarisation over the star: 
either this region is brighter than the rest, or its light is more polarised. 
This second case  of polarisation excess can also be split into two phenomena:  either light is emitted from a higher region in the atmosphere (which translates into an increased  
anisotropy $J^2_0$ of the radiation field illuminating the atoms),  or the   emitting region is moving with respect to the photosphere, the presence of a velocity gradient 
amplifying the radiation anisotropy \citep{de_kertanguy_theoretical_1998}.
As a matter of fact, these three phenomena are not exclusive and one could, as an example, call for a convection cell bringing up hot plasma, hence brighter, in a strong upflow 
(hence an anisotropy amplifying velocity gradient) that brings this plasma high in the atmosphere (hence increasing the anisotropy on its own). 

When considering these non-exclusive alternatives  we must keep in mind  the pulsation of $\chi$\,Cyg that may amplify any anisotropy present in the lower layers of the star.  And this leads us to consider that the origin of this anisotropy may be found in the non-homogeneity of the pulsation itself.  The net observed polarisation would arise from the differential amplification of anisotropy by the larger velocity gradients of those regions  expanding faster than all others, a hypothesis  already hinted by \cite{fabas_shock-induced_2011}   and  proposed by \cite{carlin_temporal_2013} for the prototypical Mira star, $o$ Cet.
In what follows we will be led, step by step, to conclude that this  scenario is the one at work in $\chi$\,Cyg.

\subsection{Asymmetry from inhomogeneous  velocity fields}

\cite{carlin_scattering_2012} have studied the amplification of the anisotropy by velocity gradients along the vertical direction. 
A remarkable result of those studies is that emission and absorption lines produce opposite effects in the anisotropy: for otherwise identical lines, an amplification of the anisotropy happens 
if the line is in absorption but a reduction in anisotropy happens if it is in emission. 
Explicitly, a solution for the intensity around a spectral line of width $w$, with continuum at $I^{(0)}$, at a certain distance of disk centre $\mu$, and 
in the presence of a linear limb darkening with coefficient $u$ can be written as:
\begin{equation}
I(\nu,\mu)=I^{(0)} (1-u+u\mu)\left[1-ae^{\frac{-(\nu-\nu_0)^2}{w^2}}\right]
\end{equation}
Assuming that this spectrum illuminates the upper atmosphere where the line presents an absorption profile of width $\Delta \nu_D$ moving at a velocity 
$\mathrm{v}_z$ along the vertical direction respect to the lower layers,  the anisotropy  in the low velocity limit can be written as in \cite{carlin_scattering_2012}
\begin{equation}
\frac{J^2_0}{J^0_0}=\frac{u}{4\sqrt{2}(2-u)}+a\frac{64-56u+7u^2}{120(2-u)^2(1+\alpha^2)(\sqrt{1+\alpha^2}-a)}\xi^2
\label{J20amp}
\end{equation}
where $\alpha=\frac{\Delta\nu_D}{w}$ and where the velocity gradient $\mathrm{v}_z$ is adimensionalized as $\xi=\frac{\mathrm{v}_z}{c}\frac{\nu_0}{w}$.
In the absence of both limb darkening and velocity gradients, the anisotropy is 0. 
In the absence of a velocity gradient, the anisotropy is due to limb darkening only.
More surprisingly, in the absence of limb darkening one can still have an anisotropic radiation field if the upper atmosphere is moving with respect to the static background, 
and this anisotropy is independent on whether the movement is upward or downward. 
Of more immediate impact in our present discussion, the sign of the anisotropy modification by velocity gradients is determined by the sign of $a$.
That is, for an absorption line ($a>0$), velocity gradients amplify any anisotropy due to limb darkening, while for an emission line ($a<0$) the effect is the opposite.

Among the different scenarios sketched above to produce net linear polarisation over the stellar disk of $\chi$\,Cyg, we can ascertain the impact of velocity gradient 
amplification by comparing emission and absorption lines in the linearly polarised spectrum. 
If in a certain region of the expanding stellar atmosphere there is a gradient of velocity larger than elsewhere, it will amplify the anisotropy of absorption lines that will increase its polarisation and dominate the net polarisation over the disk. But this larger gradient will diminish the anisotropy of emission lines which will be less polarised than elsewhere over the disk. The net polarisation in emission lines will be dominated by the regions where local polarisation is the opposite one of the region with larger gradients. Summing up, emission and absorption lines, after integration over the disk, will present opposite polarisations.

Fig.\,(\ref{CaIR}) shows the interesting case of the triplet of \ion{Ca}{II} lines in the near infrared. 
The three components show  emission in intensity and,  a small redshifted absorption, most visible at 866.2nm and difficult to ascertain in intensity in the two other lines.
Such profiles are very common in Mira stars \citep[e.g.][]{Gillet:1988aa}, and are interpreted in the framework of shockwave propagation. The emission component, blueshifted, originates at the shock front zone, propagating outward, while the red component, in absorption, comes from the unperturbed medium above the shock front. Recalling the pulsation phase of this observation ($\varphi\sim 0.94$), the spectrum has actually been obtained very close to a maximum light i.e., 
during the maximum outward acceleration  of the shock. Following our argument comparing emission and absorption lines, in agreement with Eq.\ref{J20amp}, and given the observed polarisation signs of the other lines in absorption  the  emission component of these lines is expected to show a negative $Q$ signal, while the one  in absorption   would present  a positive $Q$ signal, identical to the Na $D$, \ion{Ti}{i} or molecular lines explored above.

This  can be verified  in Fig.\,\ref{CaIR},}.
The signals are  clear in the \ion{Ca}{II} 866.2\,nm and in \ion{Ca}{II} 854.2\,nm but absent in \ion{Ca}{II} 849.8\,nm. 
This is not unexpected, in view of the particularities of the  intrinsic polarisation of these lines with a strong impact of lower level atomic polarisation 
in their statistical equilibrium and emission terms \citep{Trujillo-Bueno:2003aa,carlin_scattering_2012,carlin_chromospheric_2015}.  % {\bf il serait interessant d'avoir les $W_{ul}$ des 3 raies...}

%%%%%%% agnes  : j'ai commente les  figures car je n'ai pas les fichiers, mais les captions sont OK)

\begin{figure*}
   \centering
   \includegraphics[width=\textwidth]{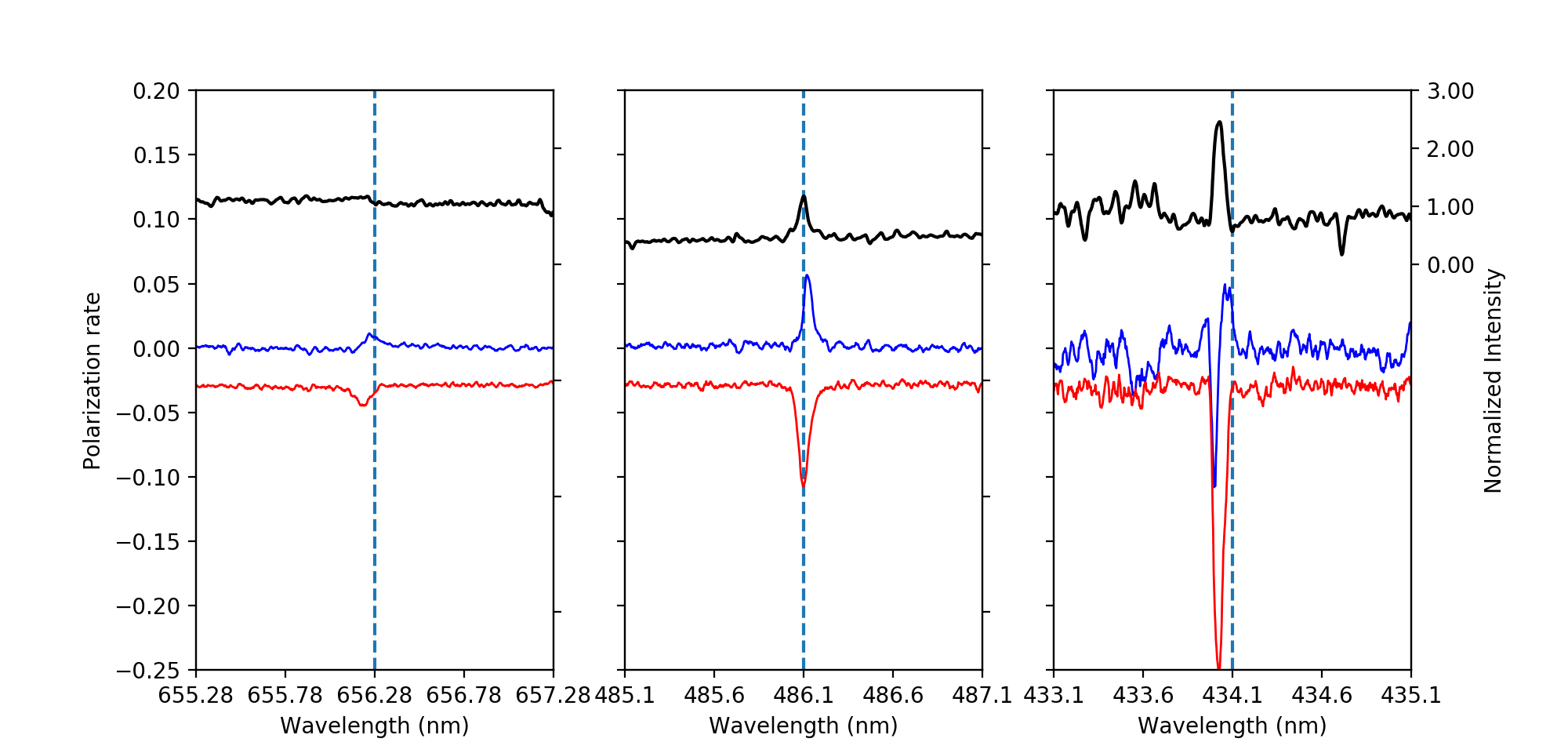}
      \caption{Spectra of $\chi$\,Cyg on September 4th, 2007 around $H\alpha$ (left), $H\beta$ (centre) and $H\gamma$ (right).
	   As in Fig.\,\ref{NaD}, the black thin line represents intensity (ordinates at right) and Stokes $Q$ and $U$ are shown in red and blue respectively. $Q$ has been shifted 0.01 units for clarity.}
         \label{Balmer}
   \end{figure*}

The effect can be cautiously confirmed by considering the $H\alpha$,  and $H\beta$  lines (Fig.\,\ref{Balmer}). 
Interpreting the polarisation of Hydrogen lines (of the Balmer series or any other) is extremely complex due to the sensitivity of this atom to all possible polarizing effects, 
and the large regions and conditions on which these lines can form \citep{lopez_ariste_full_2005,casini_scattering_2006,Casini:2006aa,Derouich:2007aa,Stepan:2008aa}, especially in a NLTE context. 
But once one accepts all our working hypothesis, that is  that most lines are showing intrinsic polarisation, and that the net polarisation is apparently due to anisotropies in the radial 
velocity of the pulsating atmosphere (amplifying the radiation anisotropy), and with the comfort that with these hypothesis we have been able to predict  the signs on the combined emission/absorption profiles of the
\ion{Ca}{ii} lines, then it is acceptable to peek into the H lines for a further confirmation. 
In the available data, $H\alpha$ does not show clear signals in intensity, perhaps because the molecular bands strongly absorb the flux originating in the corresponding line
forming region.
Conversely, the emission profile is clear in the intensity profiles of $H\beta$ and $H\gamma$, where the molecular absorption is strongly reduced.
%{\bf a verifier: Also, these lines being on the blue side of the visible spectrum, their source function depends more on the scattering terms than in the Planck emissivity.  }
The polarisation signals, on the other hand, are strong and unambiguous for the three Balmer lines. 
And, in broad agreement with the signs observed in the emission part of the \ion{Ca}{ii} lines (Fig.\,\ref{CaIR}), the emission presents a negative polarisation signal in $Q$, though we also see, particularly in $U$, a redshifted positive signal that could signal a redshifted absorption profile not visible in the intensity profiles. The complexity of formation of these lines prevents us from interpreting these signal beyond these simple facts and comparisons for these lines.
%{\color{red} {\bf Cf le papier de Fabas et al. 2011, bien que consacré à omi cet, une Mira plus froide et plus O-rich que chi Cyg, il montre toutefois des profils  U\&Q asscociés à ces raies de Balmer mais qui ont d'autres formes : émissions (aussi à la phase proche du max), et parfois émission en U et absorption en Q à la même phase. Cela exclut- il un cadre général pour les Miras, étoiles pulsantes avec choc ?}}

%%%%%%% agnes  : j'ai commente les  figures car je n'ai pas les fichiers, mais les captions sont OK)

We have thus a scenario that explains the main features of the linearly polarised spectrum of $\chi$\,Cyg. 
The particular conditions of the atmosphere of this star around its maximum brightness are such that most lines show intrinsic line polarisation. 
%{\bf un vrai test serait de verifier cela pour la meme phase d'un autre cycle... {\color{red} OUI ! et on a les moyens de le faire car dans les data de 2015-2016, on a des obs à cette meme phase ! en 2015  et en 2016!}}
This explains the signal in \ion{Na}{i}\,$D_2$ and its absence in $D_1$. 
The same signal can be expected in all lines, atomic or molecular, with an amplitude scaled by the quantum coefficient $\left(w^{(2)}_{J_uJ_l}\right)^2 $.  
We have illustrated the case with a \ion{Ti}{i} line and with 3 molecular bandheads. 
The exercise could be continued with other lines, both atomic and molecular, that present the same trends of signs and relative amplitudes in $Q$ and $U$ signals. 
Following this scenario, the pulsating atmosphere during  a maximum light has not the same radial velocity all over the star. 
Particular locations are expanding faster than others.
Since velocity gradients can amplify the radiation anisotropy of the expanding layers, the light from these locations is more polarised than elsewhere. 
Consequently, a net linear polarisation appears after integration over the stellar disk.  
But this amplification of the anisotropy and of the polarisation by these velocity gradients only happens for absorption lines. 
Emission lines actually recover a smaller anisotropy in the same conditions. 
For emission lines, the polarisation is reduced by these velocity gradients, and the net polarisation is dominated by the opposite signal coming from elsewhere in the star.
%is dominated by those regions with the smaller pulsation velocities, somewhere else in the star. 
What we expect is a change of the polarisation sign for emission lines. 
This is actually what is observed  in two of the lines of the \ion{Ca}{ii} triplet for which there is a $I$-blueshifted emission and a redshifted absorption, corresponding to a change of sign 
in the polarisation as one moves from one side to the other of the line profile. This explains also why the Balmer lines, which are mostly in emission, do show different and predominantly opposite polarisation signs to the other 
atomic and molecular lines in absorption.
This  change in the sign of polarisation  in emission and absorption lines cannot be easily explained in the two other scenarios for symmetry breaking.   If velocity gradients were identical all over the star, everywhere the local polarisation of absorption lines would be larger than that of emission lines. In that case, if one particular region was either brighter or higher than other regions, the  total amount of photons, or the polarisation rate, would be larger than elsewhere. And it would be so for  both emission and absorption lines. But the polarisation of emission lines in that brighter region would still be larger than the polarisation of emission lines elsewhere, so that the net polarisation of emission lines would be the one of that brighter region. The net polarisation of both emission and absorption lines would  show the sign corresponding to that brighter region, and both emission and absorption lines would show the same  sign even though the polarisation amplitude in  emission lines would be smaller. Not only the change of sign would not be explained, but neither do we  see  this smaller amplitude of polarisation in emission lines.

\section{Time series of LSD profiles}

Our scenario of anisotropic velocity gradients implies that all absorption lines will present the same sign of polarisation, and, excluding deep lines for which second order 
effects (lower level polarisation, hyperfine structure, ...) may be important, the spectral shape of that polarisation signature is the same. 
As already claimed above, this justifies the use of LSD techniques applied to absorption lines and weighted with the coefficient $\left(w^{(2)}_{J_uJ_l}\right)^2 $.

\begin{figure}
   \centering
   \includegraphics[width=0.5\textwidth]{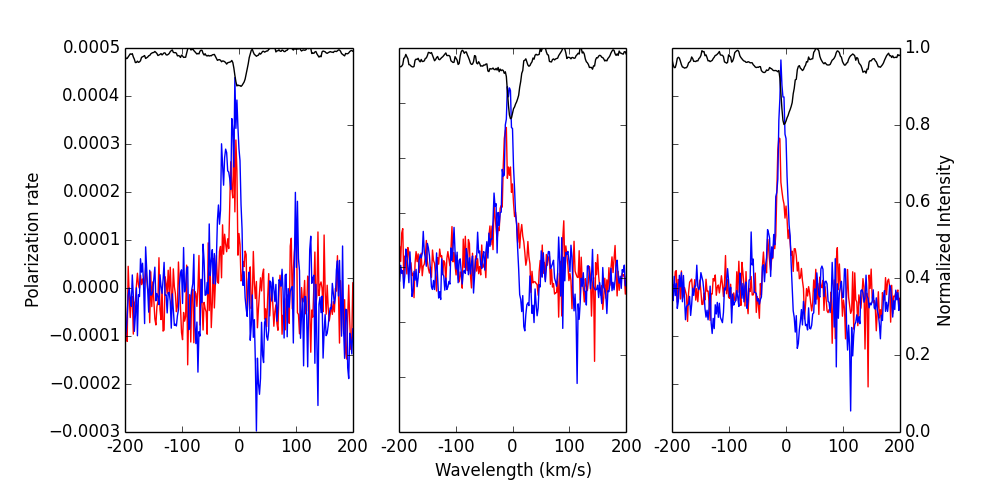}
      \caption{LSD Spectra of $\chi$\,Cyg on September 4th, 2007, with lines weighted by $(w^{(2)}_{J_uJ_l})^2$. 
	  $Q$ is in red, $U$ in blue. 
	  Three different masks are used  from left to right: all lines deeper than 1\,\%, all lines deeper than 40\,\%, and a selection of lines deeper than 40\,\% (see text). }
         \label{LSD0}
   \end{figure}

The observations of September 4th, 2007 are remarkable in that they show clear signatures in individual lines. 
For this date we can compare the LSD profile (Fig.\,\ref{LSD0}) to the signals in individual lines.
We recover the expected signature, proving that LSD really extracts  the same signals that we have interpreted. 
The LSD profile also shows a high S/N ratio, its fundamental purpose, and this demonstrates that continuum depolarisation, even if present in some lines, must be really 
a minor effect: the LSD signal is dominated by intrinsic polarisation. 
From the point of view of the Second Spectrum we find that Betelgeuse (dominated by depolarisation) and $\chi$\,Cyg (dominated by intrinsic line polarisation) place themselves 
at the two extremes afforded by Eq.\eqref{RTE}, while the Sun is to be found somewhere in between with most of the lines depolarizing the continuum and a few remarkable lines 
showing intrinsic line polarisation. 
These differences must come from the actual atmospheric structure and the formation regions of the continuum and the spectral lines. 
%{\bf je ne sais pas s'il est possible de preciser au niveau des parametres physiques, densite, taille des cellules convectives (pas pour le Soleil), etc...? {\color{red} la différence porte plutot sur la taille de l'atmosphère (extension extreme pour Mira et RSG), et sur la complexite de la dynamique pouvant dès lors s'y dérouler, non ? }}

%%%%%%% agnes  : j'ai commente les  figures car je n'ai pas les fichiers, mais les captions sont OK)
   
Signals will often be much lower at other observation dates and we are therefore interested in using the right mask of lines for LSD, right in the sense that the S/N 
ratio is still large enough but  pollution from molecular lines is as small as possible, given the lack of sufficient information to include molecular lines in the masks. Educated guesses led us to try and add all atomic lines deeper than 1\,\% (left plot of Fig.\,\ref{LSD0}), all atomic lines deeper than 40\,\% (central plot) and, right plot, 
those lines deeper than 40\,\% that belong to the following series of atomic species: 
\ion{Ti}{i}, \ion{Ti}{ii}, \ion{Sr}{i}, \ion{Sr}{ii}, \ion{Fe}{i}, \ion{Fe}{ii}, \ion{Cr}{i}, \ion{Cr}{ii}, \ion{Co}{i}, \ion{Co}{ii}, \ion{V}{i}, \ion{V}{ii}, \ion{Ni}{i}, 
\ion{Ni}{ii}, \ion{K}{i}, \ion{K}{ii}, \ion{Zr}{i}, and \ion{Zr}{ii}. 
%{\bf Donc combien de raies au total? {\color{red}   A préciser en effet pour chacun des masques.  Et aussi il faudrait sans doute justifier le choix de ces éléments ; ou la non sélection d'autres.... }}
This last LSD mask , with over 19000 lines and which results in the most clear and unpolluted signal from molecular lines without loosing much in S/N, is retained and applied in the following.  
%{\bf je me demande si comparer les regions bleues (pas trop de molecules) et rouges (beaucoup de molecules) atait visible sur les LSD? {\color{red}  en effet ce serait un bon test et  aussi un autre test : choix de fenetre(s) lambda-min-lambda-max avec/sans contrib. moléculaires}}

   %%%%%%% agnes  : j'ai commente les  figures car je n'ai pas les fichiers, mais les captions sont OK)   
    \begin{figure*}
   \centering
   \includegraphics[width=\textwidth]{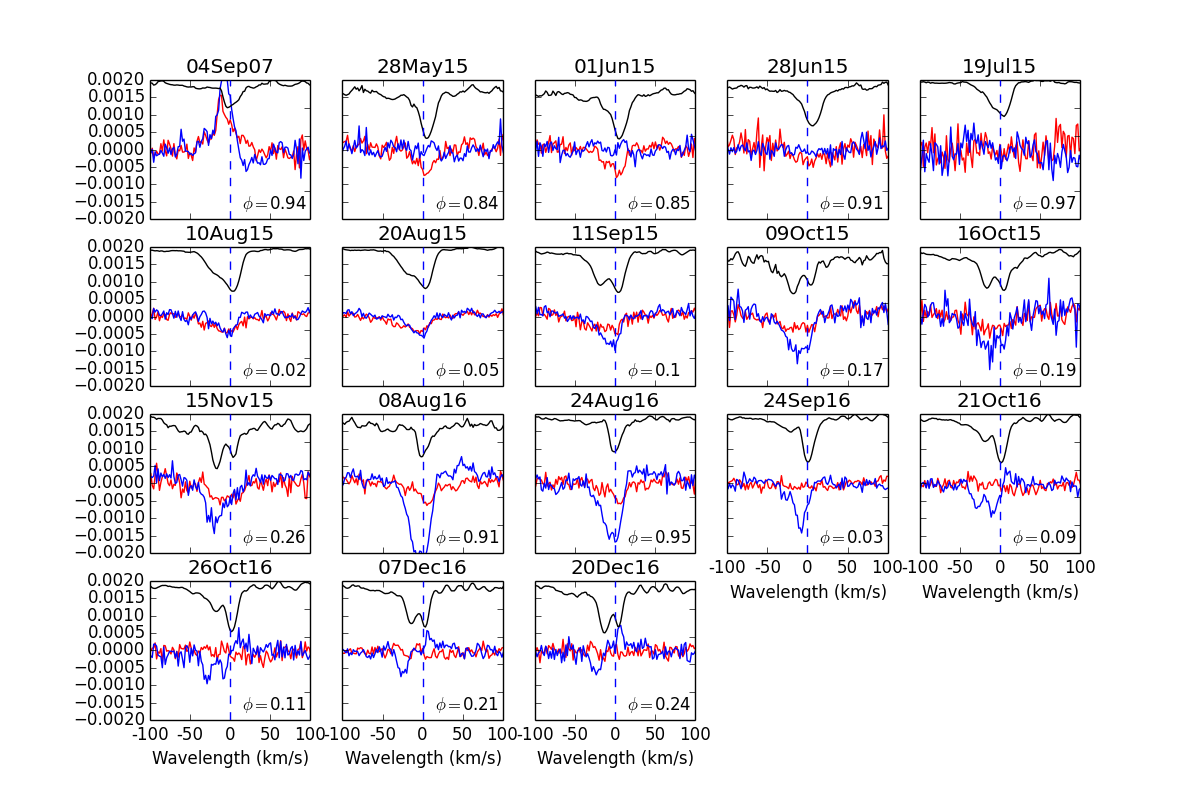}
      \caption{LSD Spectra of $\chi$\,Cyg for each one of the dates for which Stokes $Q$ and $U$ data is available. 
	    The vertical dashed line in each plot marks the heliocentric zero wavelength.  The inset text contains the phase at each observation.
	    }
         \label{LSDseries}
   \end{figure*}
   
Figure\,\ref{LSDseries} shows the LSD profiles for all the available dates. 
Signals are not always visible, as for example on June 28th, and July 19th, 2015.  % Deplacé en Section 2 : Figure\,\ref{AAVSO} shows the time series of the visual magnitude of $\chi$\,Cyg as measured by the AAVSO. The vertical dashed lines mark the dates for which LSD profiles are shown, all of them around maximum. Except the spectrum obtained in September 2007, all the other observations cluster around two consecutive maxima of the star in 2015 and 2016. 
A first point to notice is that the LSD signals change from cycle to cycle but they are coherent inside each cycle and an evolution in time can be ascertained.
We also see that, in general, the polarisation signal appears centred or blueshifted compared to the line profile minimum.
However, for the observation of December 20th, 2016, two signals with different signs appear in apparent relation with the double line in the intensity profile.
On the other hand, the peaks in $Q$ and $U$ appear at the same wavelength, up to the precision allowed by the S/N ratio. 
These observed wavelength dependences of the signal are to be understood in the basic scenario which also explains the doubling of the intensity profile, clearly visible in our data. 
A brief description was given above when explaining the emission profile of the \ion{Ca}{ii} triplet. 
The observed line doubling in intensity is  explained in the framework of the pulsation shock wave: the shockwave physically separates  the line forming regions in two with two different velocities.
The atmosphere above the shock follows a ballistic motion (induced by a former shock propagation), and produces the redshifted  feature, while the region just crossed by the shock is rising, 
leading to the blue-shifted component.
This simple description of a 1D Schwarzschild mechanism \citep{SCH52}   needs to be understood in the context of the integration over the stellar disk.  Clearly, it is only around disk centre that the two velocities result in a maximum Doppler shift visible as a double line.  As one approaches the stellar limb the projection onto the line of sight of the velocities drops to zero, cancelling any Doppler shift and eliminating the double line. If a double line is to be seen after integration over the stellar disk,  the simplistic 1-D picture of the Schwarzschild mechanism must be completed.
To address this issue, we follow \cite{bertout_line_1987} and \cite{wagenblast_spectral_1983} and we redraw their scenario for the formation of spectral lines in a moving spherical shell.  
This is explicitly done in the Appendix. 
The bottom line of their, and our, reasoning is to assume that the bulk expansion velocity is larger than the intrinsic width of the line. 
But, from the point of view of the observer, this is only true for a region around the disk centre, while near the limb the projected velocity is smaller than the line width. 
As shown in the Appendix and by \cite{bertout_line_1987} and \cite{wagenblast_spectral_1983}, the integral over the stellar disk of the emergent intensity profiles carries a geometrical factor that, under these stated  relations between bulk velocity and line width,
translates into a wavelength modulation. 
Independently of any other radiative transfer effect, this disk integration will produce doubled  profiles, shifted profiles, asymmetric profiles or flat-bottom profiles for the intensity
depending on the actual value of the expansion velocity of the shell and in the presence or absence of a contracting shell on top of the expanding shell.
  
In this very same scenario, polarisation can be computed  easily by introducing the factor $\sin^2 \theta = 1-\mu^2$ into the integration  over the stellar disk.
As shown in the Appendix, this dependence must be added to the integrals in a similar way to the geometric factor  described by  \cite{bertout_line_1987}. 
It modifies the result in  that polarisation comes preferentially from somewhere midway between the centre and the limb. 
At this position, the Doppler separation is not so large and we expect polarisation to be roughly centred in wavelength when compared with the doubled intensity profiles. 
On top of this median signal we have to consider the actual details of the position of the anisotropy in velocities which will favour the polarisation signals of 
certain places of the disk.

The interesting point in the previous description of the line formation in $\chi$\,Cyg is that  wavelength can be soundly related to a distance to the disk centre while, 
as usual, the ratio of $Q$ to $U$ provides information about polar angle. 
This conclusion leads us, as it did in the work by \cite{auriere_discovery_2016} and \cite{lopez_ariste_convective_2018}, to the possibility of actually mapping the velocity field  modulation of the pulsation over 
the disk of $\chi$\,Cyg.
   
%UNE DISCUSSION ICI SUR LES DIFFERENTS MAXIMUMS DE L'ETOILE ET LEUR SIGNIFICATION    {\bf Quels maxima? de la courbe de lumiere?}

% FIGURE DEPLACEE et PLACEE en SECT. 2   
%    \begin{figure*}
%   \centering
%   \includegraphics[width=\textwidth]{AAVSO_figure.png}
%      \caption{Visual magnitude of $\chi$\,Cyg as measured by AAVSO. 
%	    The vertical dashed lines mark the dates for which spectropolarimetric data are available. {\bf enlever la figure de gauche}}
%         \label{AAVSO}
%   \end{figure*}

%{\bf  IL SERA INTERESSANT DE PREDIRE CE QUE LA POLARISATION DEVRAIT ETRE AU MINIMUM POUR COMPARER DANS LE FUTUR.  SI LES VITESSES DE RETOUR SONT IDENTIQUES A CELLES DE DEPART, XI**2 IMPLIQUE LE MEME SIGNAL, EXCEPT POUR LA NONEXISTENCE DE RAIES EN EMISSSION (????).   MAIS SI LE PROFIL DE VITESSES DE RETOUR EST DIFFERENT ALORS LE SCENARIO S"ENRICHIT. Je tente! c'est probablement mal dit, mais il y a peut-etre des choses a creuser sur la mapping au rayon maximum.}

Even if its origin is not really known (see next section), the anisotropy of the radial velocity field is required to explain the net observed linear polarisation.
In the scenario developed above, this velocity field modulation appears at all observed phases, that are mainly centred on light maxima that correspond to the outward propagation 
of the shock wave.
In the Schwarzschild framework, this is well established by the  gradual development of a blue shifted component that generates  line doubling or asymmetric profiles in the $I$ spectra. 

However, once the shock wave has crossed the atmosphere, the accelerated matter follows a classical ballistic motion, reaching, around light minimum, its maximum extension.  
It would be interesting to measure the linear polarisation at this  phase, since we do not expect strong inhomogeneities within the velocity field in the line forming region at this phase of maximum extension.  Trials to do so have unfortunately been hindered by the even more ubiquitous molecular bands that prevent  the computation of any meaningful LSD profile using masks of atomic lines exclusively.
% {\color{red} {\bf  ben justement on a des data à ces phases là !  Au minimum, on ne peut plus tracer la présence du choc dans l'atmosphère et l'atmosphère se relaxe (mvt descendant) donc on s'attend à voir en I un profil  redshifté. pb au minimum de lumière, l'atmosphère est très froide, le SN des obs de phi = 0.3 à phi =0.7 va être faible et le profil des raies fortement mutilé par des absorptions moléculaires.   }} 
%In addition, since spherical symmetry is not broken,  any linear polarization signal  present must only be due to brightness inhomogeneities (Section 3.3), that may be
%related to convective cells.
%If the corresponding mapping traces the velocity anisotropy of the precedent (or next) phase, then radiative pressure may be at its origin, if it is out of phase, the velocity anisotropy
%might be linked to that associated to the cells hydrodynamics. {\color{red} {\bf  La fin de cette dernière phrase est incomprehensible pour moi ; merci de  la reformuler !}}

\section{Asymmetries in the shock wave}
The main conclusion of this work has by now been reached: the shock wave in $\chi$\,Cyg should present asymmetries in the form of an inhomogenous radial velocity field. 
In order to quantify those asymmetries, we can attempt to map them onto the stellar disk using the conclusions from the analysis presented in the previous section on the formation 
of the polarised line.  
This cannot be done, however, without assuming a long list of hypotheses and approximations that we try now to make explicit. 
Seen together, all those hypothesis and approximations imply that the produced maps cannot be seen at this point as the actual representation, point by point, 
of the actual velocity field in the pulsation wave of $\chi$\,Cyg. 
But these maps can be a source of information on spatial and temporal scales of the inhomogeneities, as well as their relative importance with respect to the average velocity 
of the shock. 

With this caution in mind, we start by recalling that the ratio $Q$/$U$ provides the polar angle $\chi$ of regions over the disk that contribute with a larger polarisation 
because of a larger velocity field. 
But they do so through the expression
$$\tan 2\chi =\frac{U}{Q},$$
so that a 180 degrees ambiguity pops up: the emitting region of the polarisation excess may be at one side or the other of a given diameter across the stellar disk. 
In our maps we will select arbitrarily just one of the two possibilities.
 
In our model to explain the observed linear polarisation profiles, we have assumed that the only source of linear polarisation is scattering and that no phenomena other than 
velocity gradients modifies this polarisation. 
But it is well known that magnetic fields, through the Hanle effect, can both diminish this scattering polarisation and rotate the polarisation plane \citep[e.g.]{landi_deglinnocenti_polarization_2004}
% {\color{red} {\bf  Et donc ? la présence d'un faible champ mag tel que celui détecté à la surface de chi Cyg a t elle un tel impact ?? L'effet Hanle est sensible a des champs plus faibles que l'effet Zeeman. La question est donc si les mesures Zeeman sont faibles parce que le champ est faible (donc susceptible d'induire effet Hanle) ou si elles sont faibles parce qu'il y a melange de polarites}}
Less known is that non-vertical velocity fields can produce the same effect as the Hanle effect: that is, it can depolarize and rotate the polarisation plane \citep{de_kertanguy_theoretical_1998,landi_deglinnocenti_polarization_2004}.
This rotation of the local polarisation plane can mask the dependence of the $Q/U$ ratio on the polar angle. 
So we are making the hypothesis here that the velocity field is strictly radial and that no magnetic fields are present, this last approximation made in spite of the 
measurement of such fields through the observation of the Zeeman effect in the circular polarisation of the lines \citep{fabas_shock-induced_2011,lebre_search_2014}. %{\color{red} {\bf  C'est donc un peu violent comme hypothese non ? }}. 
While imposing a radial velocity field in a scenario of a radially pulsating star may appear as a natural hypothesis, overlooking the impact of actually measured magnetic fields may be judged inappropriate.  It can be argued however that Zeeman effect is sensitive to fields much larger than the Hanle effect, so that those fields measured through the Zeeman effect  are most probably relatively strong fields (above 100G) which do not influence scattering  linear polarisation. If the measured circular polarisation amplitudes are small (1-2 G) it is just because of polarity mixing or dilution in a non-magnetic atmosphere, and not because of intrinsically weak fields. 
 
We are also assuming that at any time the star is  spherical. 
If it were an ellipsoid (as huge rotational speeds in other stars appear to do) there would be a natural source of net polarisation due to this shape. $\chi$\,Cyg is not a rapid rotator and if, nevertheless, this was the origin of the observed net polarisation
we would have failed in interpreting the different signs of the polarisation in absorption and emission lines, as we have done. 
So this must be at most a second order contribution.

\begin{figure}
   \centering
   \includegraphics[width=0.5\textwidth]{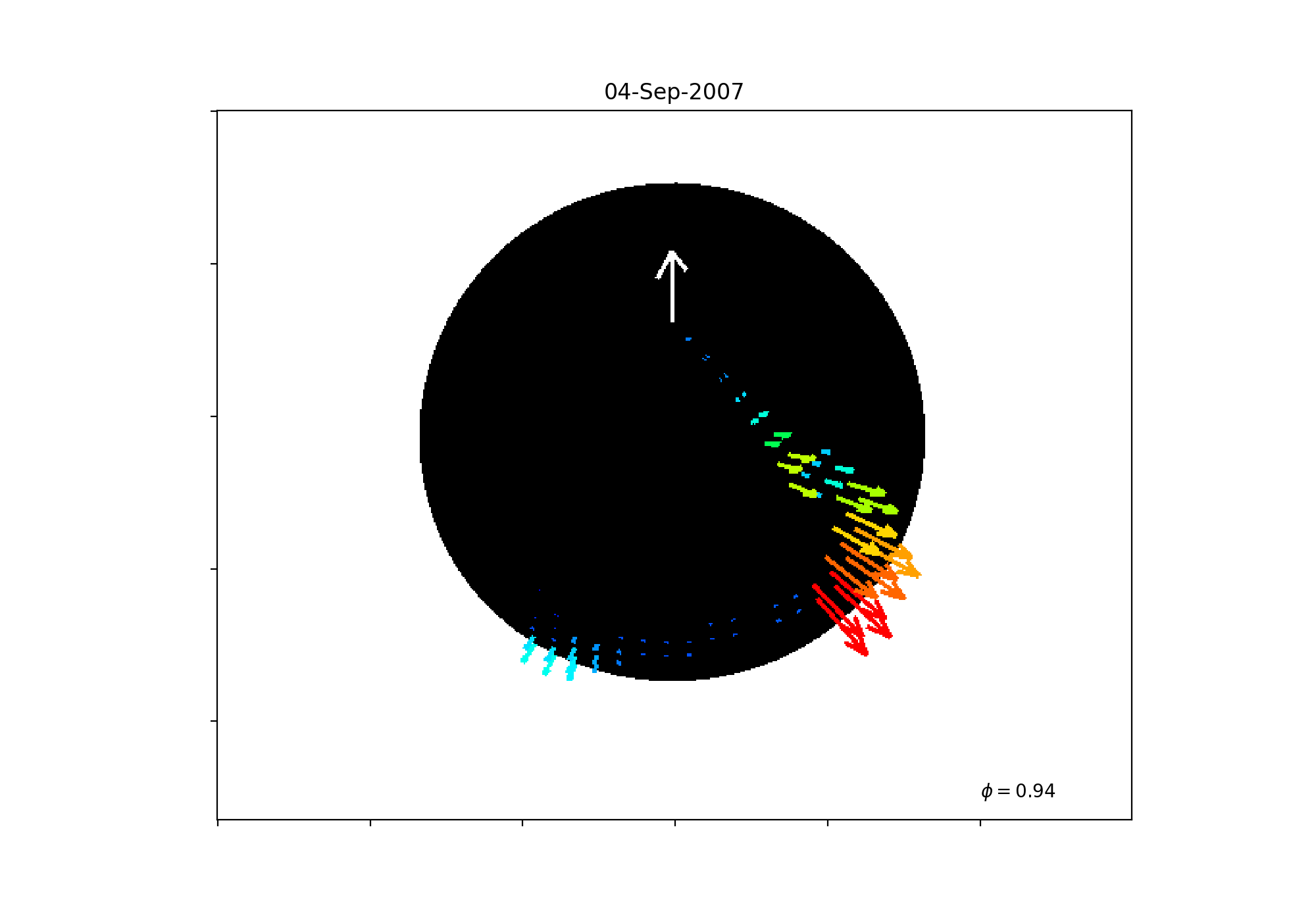}
      \caption{Map of the velocity field on $\chi$\,Cyg on Sept, 4th, 2007. A uniform radial velocity is assumed except for those places where polarisation indicates larger velocities. Both the arrow length and the colour code indicate redundantly the magnitude of the velocity in terms of the background velocity field for those places where this velocity is different. The white arrow represents the line of sight.}          \label{map4}
   \end{figure}

Another approximation comes when inferring the distance $\mu$ to the disk centre. 
Our approach has been that of the simplified model used for the imaging of Betelgeuse made by \cite{auriere_discovery_2016}: each wavelength in the profile corresponds to a unique point 
on the stellar disk of $\chi$\,Cyg. 
Thus, the recovered information concerns just as many points as there are in the spectral binning of our profiles, about 30. 
This is obviously a very rough approximation. 
In the case of Betelgeuse, it has been demonstrated \citep{lopez_ariste_convective_2018} that such simple modelling can provide, at best, only basic information of the brightest points 
(or the larger velocity points in our case). 
But it can also be the source of strange unphysical distributions. 
For example Fig.\,\ref{map4}, in which the velocity has been represented as arrows over the visible hemisphere of the star, shows strange filaments of high velocities over the disk that look similar in shape to the examples shown by \cite{lopez_ariste_convective_2018} and which have 
their origin in the drastic identification of one wavelength, one point. 
The right approach would have been to propose a model for the velocity distribution over the stellar disk in terms of spherical harmonics or others. 
However at this point we miss the broader view of what such distribution may look like so we have preferred the basic approximation of one wavelength one point from 
which at least we can retrieve some information on the largest velocity inhomogeneities present.

   %%%%%%% agnes  : j'ai commente les  figures car je n'ai pas les fichiers, mais les captions sont OK)  
  \begin{figure*}
   \centering
   \includegraphics[width=\textwidth]{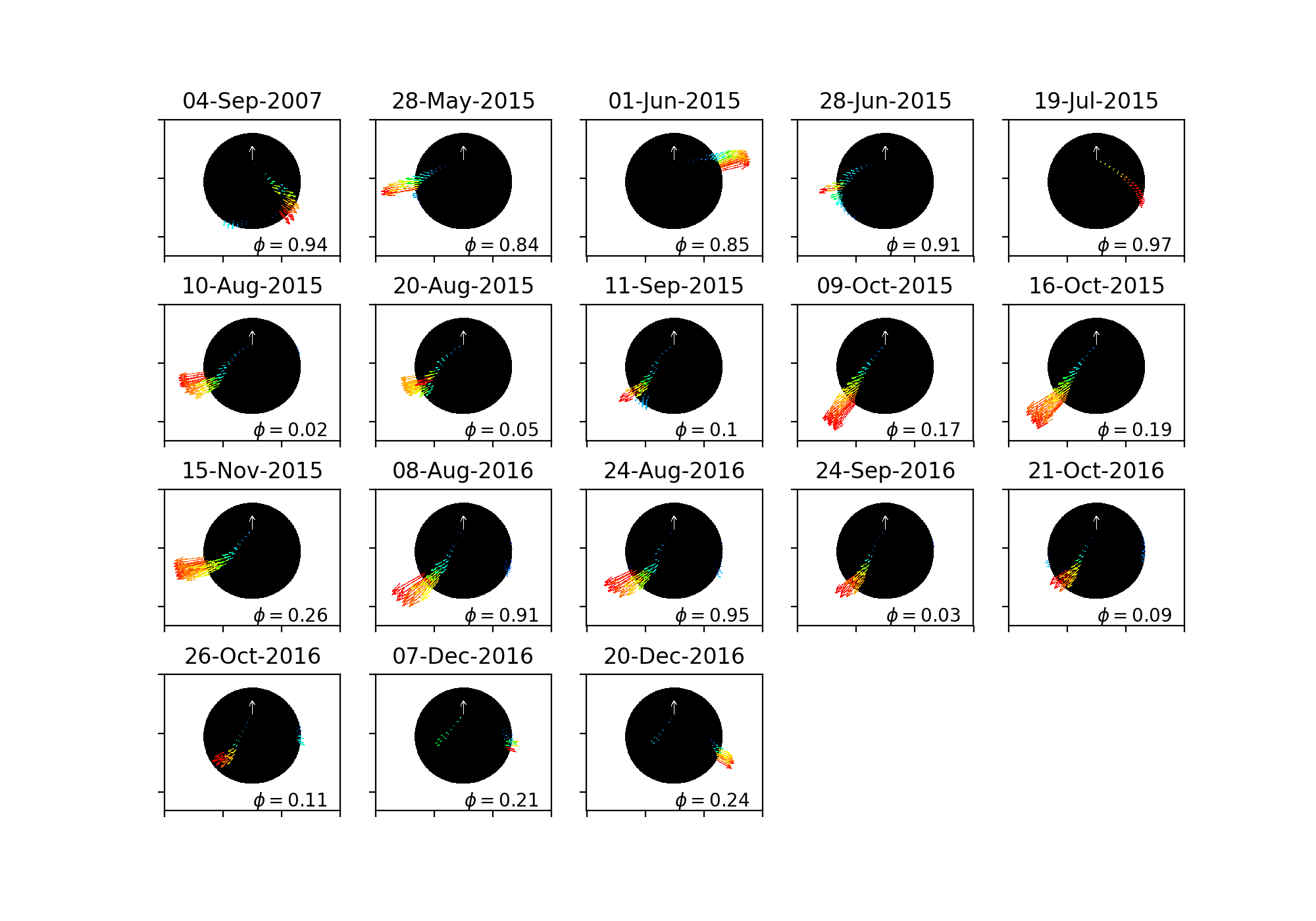}
      \caption{Maps of the velocity field on $\chi$\,Cyg. A uniform radial velocity is assumed except for those places where polarisation indicates larger velocities. Both the arrow length and the colour code indicate redundantly the magnitude of the velocity in terms of the background velocity field for those places where this velocity is different. The white arrow represents the line of sight, and the inset text gives the pulsation phase}          \label{maps}
   \end{figure*}

With these caveats in mind, Fig.\,\ref{maps} shows the inferred maps of the pulsation velocity through the sequence of observation dates with the same representation as Fig.\,\ref{map4}.  To build each of them we have gone over the $Q$ and $U$ profiles of each date and assign a distance to disk centre $\mu$ to every wavelength assuming a constant velocity projected onto the line of sight over the whole disk. This is simplistic and oversees our conclusion about velocity gradients but, once again, our goal here is to get a first picture of the spatial and temporal scales of those gradients. We next assign an azimuthal angle $\chi$ determined from the ratio of $Q$ to $U$ amplitudes at that wavelength. At that point $(\mu,\chi)$ over the stellar disk we draw an arrow proportional to the amplitude of polarisation $\sqrt{(Q^2+U^2)}$ and representing a qualitative measurement of the gradients of the velocity at that point. Since the 3D projection may make it difficult to ascertain the actual length of every arrow, we have also coded this information in colour.
As expected from the coherency of the signals, the direction of larger velocities is approximately constant around each maximum. 
The maxima of 2015 and 2016 also appear to share a common direction, but this direction is orthogonal to the one seen in the maximum of 2007. 
This would exclude a permanent preference of direction related, for example, to the rotation axis of the star.
There is also no particular correlation between the date of the maximum brightness and the date of maximum polarisation.
%{\bf je ne comprends pas finalement pourquoi l'exces de vitesse serait aussi "etroit"... quel mecanisme peut faire que le long d'un rayon sur le disk on ait un ensemble de vitesses maximales? je prefere les 2 dernieres maps que les autres...
%
%ALA:  Je crois que la forme allongee est un defaut de l'inversion. C'est comme les images de mu Ceph montrees dans la Fig.6 de l'article de 2018.}

\section{Conclusion} 
The observation of net linear polarisation in the spectral lines of $\chi$\,Cyg around three different pulsation maxima has led us to conclude that the velocity of the pulsation, 
while still radial, is not homogeneous in velocity over the stellar disk. 
The source of these inhomogeneities is not known but since they change in shape and position from one maximum to the next it must be concluded that the source is not a 
permanent feature of the star. It is nevertheless a recurrent feature, cycle after cycle, so we must also conclude that it is somehow related to the shock associated to the pulsation.

We can speculate on the interaction between the convective motions and this pulsation shock as a possible culprit: strong localized convective upflows would add up velocities with 
the  shock, resulting in localized higher velocities of pulsation. Another hypothetical source of asymmetries in the velocities would be the interaction of the shock with returning material from a previous pulsation cycle. 
These are the questions that future work will have to explore.
%{\bf si possible, il faudrait faire des maps a partir des signaux des raies NaD2 et TiI, eventuellement Hbeta, pour voir le mapping de vitesses associees a des regions tres differentes {\color{red} oui ce serait bien,  Hdelta aussi  serait un bon traceur. ALA:  Je ne sais pas interpreter les signaux des raies de Balmer parce que trop complexes. A peine si on remarque qu'elles respectent le signe de la polar. Je n'ose pas aller plus loin. Pareil avec le NaD, assigner une region de formation a cette raie me parait trop dangereux, et encore moins dire que l'intensite et  la polar se forment aux memes lieux. Il nous resterait la raie de TiI....}}

Inferring that an inhomogeneous velocity field is the cause of the observed linear polarisation signal has required the interpretation of this signal as due to the intrinsic 
polarisation of each individual spectral line. 
Disentangling intrinsic polarisation from depolarisation of the continuum, the two terms appearing in the equation for the transfer of linear polarisation in the presence of 
scattering, has been done through the inspection of individual lines. 
But we have also demonstrated that line addition, through techniques like LSD or others, can help to disentangle the two processes as well by comparing the net signal 
recovered when different weights for each individual line are used. 
Line depolarisation signatures are enhanced when the line depression is used as weight, while intrinsic polarisation is brought up by using the quantum $w^{(2)}_{J_u J_l}$ coefficient
of the line. 
This protocol allows to conclude on the origin of observed linear polarisation even in the case where the polarisation signal on individual lines is too weak. 
%It also allows to produce two LSD pseudo-lines, one for the depolarization signature another for the intrinsic polarzsation, from one single observation of stars which, 
%like the Sun, have a mixture of lines whose polarization is dominated by one or the other source. {\color{red} {\bf bof je trouve pas que cette dernière phrase soit essentielle, perso je la supprimerai. }}
%
\bibliographystyle{aa}
\bibliography{art64}
\begin{acknowledgements}
This work was supported by the "Programme National de Physique Stellaire" (PNPS) of CNRS/INSU co-funded by CEA and CNES.

\end{acknowledgements}

\appendix

\section{Integration over the disk of intensity and polarisation line profiles from moving spherical shells}
\begin{figure}
   \centering
   \includegraphics[width=0.55\textwidth]{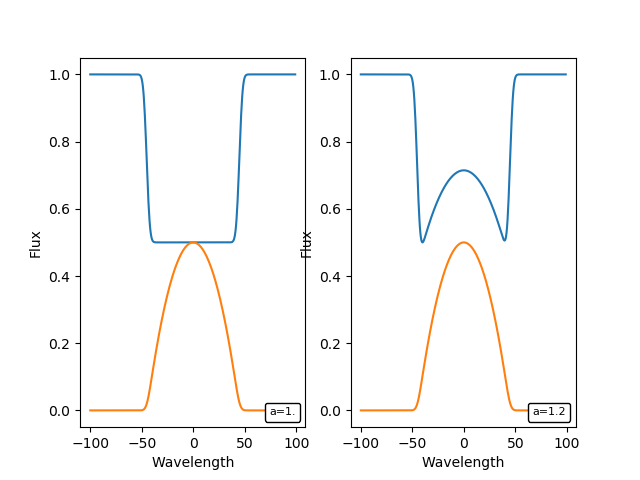}
      \caption{Examples of line formation following Bertout \& Magnan (1987). 
	  At left, a basic Schwarzschild mechanism produces a flat bottom profile captured by setting $a=1$. 
	  At right, two accelerating spherical shells with acceleration $a=1.2$, one expanding the second falling, producing the line doubling.  
	  The maximum contribution comes from the centre of the disk, the limbs being weighed down by the geometry of the problem, so the profile roughly behaves as if 
	  coming from one-dimensional radiative transfer at disk centre.
	  The red profiles correspond to the linear polarisation one.}
         \label{Bertout1}
   \end{figure}

\begin{figure}
   \centering
   \includegraphics[width=0.55\textwidth]{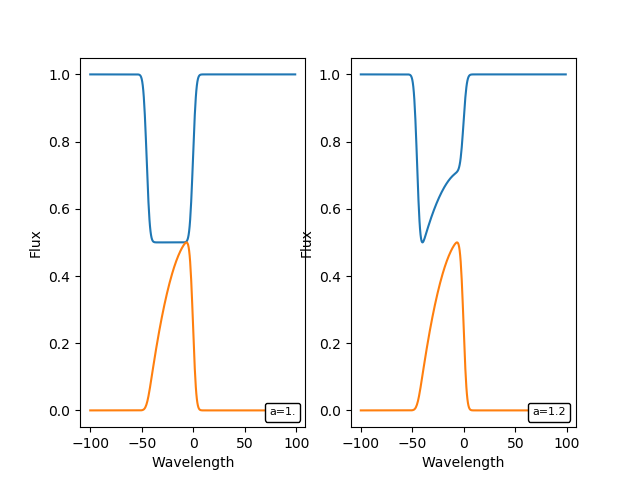}
      \caption{As in Fig.\,A.2, examples of line formation following Bertout \& Magnan (1987). 
	At left, a basic Schwarzschild mechanism produces a flat bottom profile captured by setting $a=1$. 
	At right, two accelerating spherical shells with acceleration $a=1.2$, one expanding the second falling, producing line doubling.  
	The maximum contribution comes from the centre of the disk, the limbs being weighed down by the geometry of the problem, so the profile roughly 
	behaves as if coming from one-dimensional radiative transfer at disk centre.
	The red profiles correspond to the linear polarisation one.}
         \label{Bertout2}
   \end{figure}

The doubling of the intensity profiles and, by extension, the polarisation profiles observed in $\chi$\,Cyg is attributed to the presence of one or more 
spherical shells in movement. 
The Schwarzschild mechanism \citep{SCH52} is then usually recalled \citep{alvarez_envelope_2000} to explain that, at disk centre, each moving layer will produce an individual line Doppler shifted by 
its velocity: if one layer is moving outward and another layer is falling back, two lines will be produced. 
Cartoons of this mechanism are given by  \cite{alvarez_envelope_2000}.
None of those cited works however puts any attention to the fact that, as said above, this justifies only the profile that would be observed at disk centre. 
Assuming that the velocities of the moving shells are radial and equal to $+\rm v$ and $-\rm v$,  the two lines will be Doppler shifted at disk centre by $2\rm v$, but at the limb, 
the null projection over the line of sight will produce the addition of the two lines without any frequency shift. By integration over the disk, all cases of line doubling from 0 at the limb to $2\rm v$ at disk centre will appear at different proportions and produce bottom flattened profiles (see as  examples the left profiles in 
Figs.\,\ref{Bertout1} and \ref{Bertout2} later).
Since these are not the observed profiles, we must conclude that the integration over the disk is a critical aspect of the line doubling and it is confounding that so many works ignore it.

polarisation is produced preferentially at high scattering angles. 
Thus, it is close to the limb that one expects most of the polarisation to come from. 
Following the previous scheme, at the limb the projected velocities of the moving shells onto the line of sight are zero, and the largest amount of polarisation must be emitted at 
an unshifted wavelength. 
If such a statement were true, one would expect no line doubling in the polarisation profile, but rather a well centred linear polarisation at the zero rest velocity of the star. 
This is not what is observed either.

In order to understand the polarisation line profiles we need to understand how to correctly integrate over the stellar disk and, as a particular case, how the intensity line 
profiles result in a doubled profile after disk integration. 
The best explanation of how this integration is done is found in the work of \cite{bertout_line_1987}  and \cite{wagenblast_spectral_1983}. 
Those authors already claim in their respective introductions that their purpose is, indeed, to explain the line doubling observed in Mira stars. 
Since later literature on the spectra of Mira stars has plainly ignored those works, and since they are a critical part to explain how the 1D Schwarzschild mechanism 
can still be seen as a cartoon explanation of what happens even after disk integration, and since we needed to reproduce those computations while introducing a 
$\sin^2 \mu$ factor in the integrals to explain the line doubling in linear polarisation, for all these reasons we have found interesting to reproduce in this Appendix 
the arguments and cases of \cite{bertout_line_1987}.

\begin{figure}
   \centering
   \includegraphics[width=0.5\textwidth]{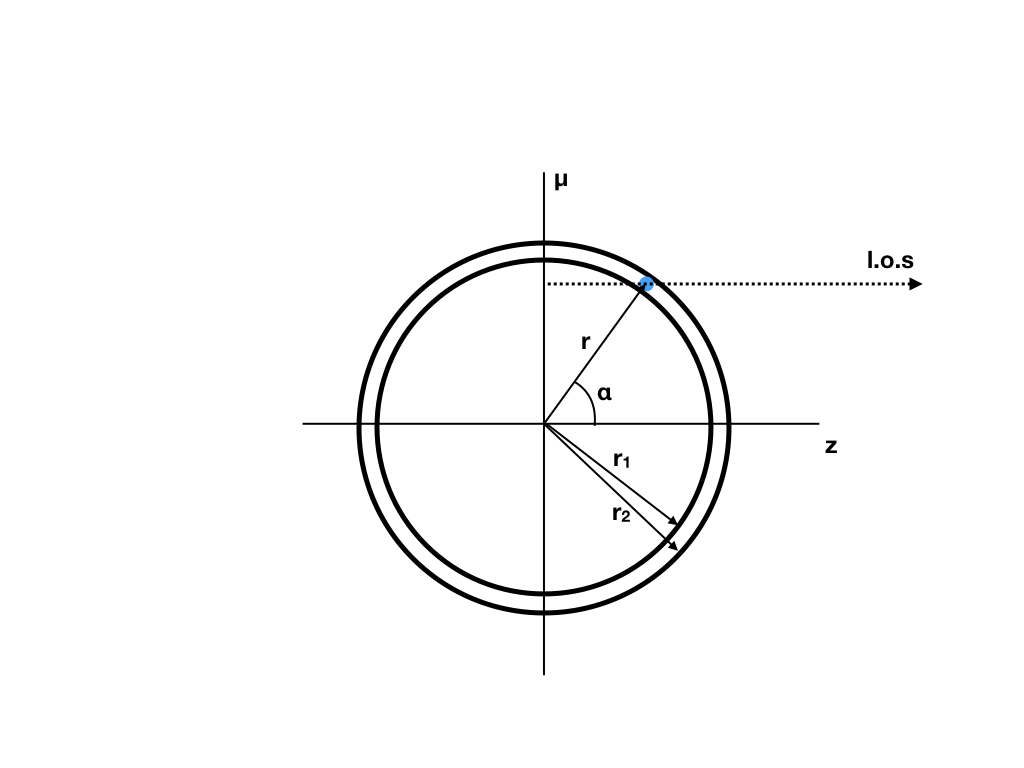}
      \caption{Geometry of the expanding spherical shell.}
         \label{cartoon}
   \end{figure}

The star is assumed to produce a continuum spectrum which is then absorbed by an optically thick spherical layer. 
\cite{bertout_line_1987} also explore the cases of optically thin layers, but we go straight into the case of interest to us and illustrate exclusively the optically thick case. 
The spherical layer is assumed to move with a velocity ${\rm v}(r)$ which depends on the distance $r$ to the centre of the star (see Fig.\,\ref{cartoon} for the graphical 
definition of $r$ and other parameters of interest).  
The layer itself has a thickness given by the difference between the inner and outer radii $r_1$ and $r_2$. 
We observe the star along a given direction $z$. 
The transverse distance $\mu$ has been defined in the main text as the distance to the centre of the disk in the plane perpendicular to $z$. 
A point in a given line of sight at $\mu$ is defined by coordinates $\mu$ and $z$, but also by a radius $r$ and an angle $\alpha$. 
Obviously we have that
\begin{equation}
r^2=\mu^2+z^2
\end{equation}
and
\begin{equation}
\cos \alpha=\frac{z}{r}=\frac{z}{\sqrt{\mu^2+z^2}}.
\end{equation}
Along a given line of sight $\mu$, radiative transfer requires the solution of an integral for the specific intensity at a given frequency $\nu$:
\begin{equation}
I_{\nu}(p)=\int S[r]e^{-\tau_{\nu}(p,z)}d\tau_{\nu}.
\label{rte0}
\end{equation}
where $S$ is the source function and $\tau_v$ is the optical opacity at frequency $\nu$, which we can write as
\begin{equation}
d\tau_{\nu}=-k \phi \left( \nu-\nu_0\frac{\rm v}{c}\cos \alpha - \nu_0 \right) dz.
\end{equation}
Symbols in this expression have their usual meanings: $k$ is an absorption coefficient and $\phi$ is the line profile centred at $\nu_0$ except for Doppler shifts.  

One of the main arguments of \cite{bertout_line_1987} is that the line doubling is a purely geometrical effect, and this cannot be better demonstrated than by simplifying radiative 
transfer to its bare fundamentals. 
Therefore, and because of our assumption of an optically thick layer, we can just impose that, independently of $\mu$, the specific intensity is just an appropriately 
defined average $<S>$ of the source function, Doppler shifted to the frequency $\nu_0\frac{\rm v}{c}\cos \alpha$. 
Correct radiative transfer calculations, as done by  \cite{fokin_hydrogen_1991} will introduce an improved line profile shape and correct line depressions. 
But the basic line doubling in the presence of a moving spherical shell can be retrieved in this bare scenario.  
With an infinitely thin line profile, for a given $\mu$, the frequency $\nu$ at which $<S>$ is non-zero is given just by
\begin{equation}
\nu -\nu_0 -\nu_0\frac{{\rm v}}{c}\cos \alpha=0.
\end{equation}
We can rewrite this condition in terms of $\mu$ by developing $\cos \alpha$:
\begin{equation}
\mu^2=r^2\left[ 1-\left( \frac{c}{\nu_0 {\rm v}}\right)^2 (\nu-\nu_0)^2 \right]
\end{equation}

From it, the contribution of a differential element $d\mu$ over the disk can be obtained by differentiation. 
Recalling that $\rm v$ is a function of $r$ and making use of the fact that
\begin{equation}
\frac{r}{\rm v}\frac{d{\rm v}}{dr} =\frac{d \log {\rm v}}{d \log r}=a
\end{equation}
we can write this differential contribution as
\begin{equation}
\mu d\mu = 2rdr \left[1-(1-a)\left(\frac{c}{\nu _0 {\rm v}}\right)^2(\nu-\nu _0)^2\right]+r^2\left(\frac{c}{\nu _0{\rm v}}\right)^2 (\nu-\nu _0)d\nu.
\end{equation}
Our calculations up to this point have transformed the original integration over opacity $d\tau$ into an integration over the disk which reduces to an integration 
over the radial coordinate $d\mu$ and then, because of the relationship between the position over the disk and the Doppler velocity of the expanding or contracting shells, 
into the sum of an integration over spectral frequency $d\nu$ and a second one through the thickness of the shell $dr$ with a strong weight favouring profiles emerging 
from disk centre over those from the limb as we shall now see.

Two different cases can be considered in view of these two terms at the right side of this equation: either the velocity of the spherical shell is such that macroscopic 
Doppler shifts are much larger than the width of the spectral line, and in such case we can neglect the term in $d\nu$, or the line is broader than the Doppler shifts due 
to the moving shell and it is the first term in $dr$ that we can neglect. 
This second case produce what \cite{bertout_line_1987} referred to as \emph{sawtooth profiles} which they considered as the probable description of the line doubled lines in Mira stars. 
However, posterior works proved that the shocks of $\chi$\,Cyg and other Mira stars present velocities with Doppler shifts much larger than the line thermal width. 
These works force us to be in the first case above, where we can approximate
$$\mu d\mu = 2rdr \left[1-(1-a)\left(\frac{c}{\nu _0 {\rm v}}\right)^2(\nu-\nu _0)^2\right].$$

This expression allows us to easily write and compute the disk integration of the emergent profiles. 
The observed flux at frequency $\nu$ will be given by the integral
\begin{eqnarray}
F_\nu=\int_{0}^{2\pi} d\chi \int_0^r I_{\nu}(p) pdp= \\
 2\pi <S> \int_0^r  rdr \left[1-(1-a)\left(\frac{c}{\nu _0 {\rm v}}\right)^2(\nu-\nu _0)^2\right]
 \end{eqnarray}
 
Let us finally assume that the shock layer is thin enough so that we can make $v$ constant through the layer. 
This allows us to write the final and simple expression:
\begin{equation}
F_\nu=\pi <S> \left[1-(1-a)\left(\frac{c}{\nu _0 {\rm v}}\right)^2(\nu-\nu _0)^2\right] (r_2^2-r_1^2).
\label{rte1}
\end{equation}

The frequency dependence of $F_{\nu}$ comes exclusively from the equations that describe $dr$ in terms of $d\mu$. 
That is, the observed flux presents a shape in frequency which is determined not by radiative transfer but by a geometric factor alone. 
Depending on the value of $a$, the logarithmic gradient of the velocity, we can find two interesting results. 
If $a=1$, that is, if the layer does not move or if it moves with a velocity proportional to $r$, then
\begin{equation}
F=\pi <S> (r_2^2-r_1^2)
\end{equation}
independently of frequency. 
This a bottom-flat square profile where no line doubling can be seen. 
This is the case illustrated in the left plots of figures\,\ref{Bertout1} and \ref{Bertout2}, and shows what a straightforward disk integration of the 1D Schwarzschild mechanism would produce.
On the other hand, if $a>1$ (the layer is expanding in an accelerated motion), then $F_{\nu}$ has a parabolic shape where the disk centre contributes to the blue edge of the 
line with weight one, and points further and further from disk centre contribute at frequencies nearer and nearer to the central frequency $\nu_0$ but with a diminishing factor 
$(\nu-\nu _0)^2$ multiplying those contributions. 
The result is a line which grows towards its blue edge and leaves the line centre with less photons. 
A second layer, falling into the star from a previous pulsation, produces the same profile red-shifted. 
This is the line doubling observed in $\chi$\,Cyg explained exclusively by a geometric effect.  
Since the maximum depth of the line is attained at the maximum of $(\nu-\nu _0)^2$, that is from the contributions from the disk centre where the expansion velocity 
coincides with the Doppler shift, we recover a scenario where the 1D Schwarzschild mechanism correctly guesses where the doubled lines will be found, even if it misses 
completely the meaning of those doubled profiles.

In our observations the doubled line is not always apparent, but the intensity profile appears to be displaced from the line centre. 
It is sufficient to eliminate the falling layer from the computations and assume that only the expanding layer is emitting light. 
We illustrate this case in Fig.\,\ref{Bertout2}. 
The line appears here both displaced and asymmetric.
%{\bf dans l'interpretation "classique", il y a quand meme les 2 composantes qui sont presentes, censees produire un "doubling non resolu"... la tu
%dis qu'une couche en acceleration produit un profil asymmetrique (et je suis d'accord), mais ca ne veut pas dire que profil asymmetrique signifie juste que l'on ne voit qu'une couche?
%
%ALA: Je n'ai pas compris. Mais si par modele classique, tu veux dire le "mecanisme de Schwarzschild" 1d....il ne predit pas le profil correcte en polarisation ni les 2 cas possibles en fonction de l'acceleration.}

These simple calculations confirm one of the most important assumptions in our work, that every frequency of the double line corresponds to a Doppler shift coming from the 
projection of the velocity $\rm v$ onto the line of sight $\frac{\rm v}{c}\cos \mu$. 
This justifies our assignment of every frequency to one distance to the disk centre $\mu$, opening the path to mapping the velocity inhomegeneities over the disk.

The same computations can be now repeated for the polarisation by just introducing the $\sin^2 \mu$ factor into 
\eqref{rte0}:
\begin{equation}
Q_{\nu}(p)=\int S[r]\sin^2 \mu e^{-\tau_{\nu}(p,z)}d\tau_{\nu}.
\end{equation}
In the intensity case, geometry alone enhanced light coming from disk centre. 
We expect that the $\sin^2 \mu$ factor will take the geometrical effect of the expanding layer  into account by enhancing light coming from the limb. 
After rewriting this new factor in terms of the line frequencies and the coordinate $r$ using exactly the same conditions, approximations and developments of the intensity 
case, we find that \eqref{rte1} becomes:
\begin{eqnarray}
F^Q_\nu&=&\pi <S> \left[1-\left(\frac{c}{\nu _0 {\rm v}}\right)^2(\nu-\nu _0)^2\right] \times \nonumber \\
&\times&\left[1-(1-a)\left(\frac{c}{\nu _0 {\rm v}}\right)^2(\nu-\nu _0)^2\right] (r_2^2-r_1^2).
\end{eqnarray}
The new factor in that expression is identical except for the absence of the $(1-a)$ factor. 
The red lines in Figs.\,\ref{Bertout1} and \ref{Bertout2} show that the expected polarisations in the two cases of accelerated expansion and in the presence or absence 
of a falling shell. 
What we see is that the polarisation profile is always better centred than the intensity profile. 
In the case of the presence of both the expanding and falling shells, the polarisation profile is symmetric and centred, not presenting the double line that the intensity profile shows. 
When just the expanding shell is present, the maximum polarisation emission is expected very close to the central frequency, unlike the intensity profile which is shifted.  
In both cases, there is a clear correspondence between wavelength and distance to disk centre.

%Figure 1

%  \begin{figure*}
%   \centering
%   \includegraphics[width=0.45\textwidth]{}
%  
%   
%      \caption{}
%         \label{}
%   \end{figure*}
%

\end{document}